\documentclass[twocolumn,10pt]{IEEEtran}

\usepackage{amsmath}
\usepackage{mathtools}
\usepackage{enumerate}
\usepackage{paralist}
\usepackage{graphicx}

\graphicspath{{Figures/}}
\usepackage{subfigure}
\usepackage{cite}
\usepackage{array}
\usepackage{booktabs}
\usepackage{comment}
\usepackage{cases}
\usepackage{algorithm}
\usepackage{algorithmic}
\usepackage[mathscr]{euscript} 
\usepackage{url}
\usepackage{amssymb}
\usepackage{amsthm} 

\usepackage{etoolbox}

\usepackage{color}

\usepackage{multirow}
\usepackage{array}
\usepackage{ragged2e}
\newcolumntype{P}[1]{>{\RaggedRight\hspace{0pt}}p{#1}}
\newcolumntype{C}[1]{>{\centering\hspace{0pt}}p{#1}}

\usepackage{epstopdf}

\newtheorem{theorem}{Theorem}

\newtheorem{proposition}[theorem]{Proposition}

\newtheorem{corollary}{Corollary}
\newtheorem{definition}{Definition}

\newcounter{myoptimizationproblemctr}

\makeatletter
\newcommand{\Vast}{\bBigg@{4}}
\makeatother

\begin{document}

\title{Distributed Resource Allocation for Relay-Aided Device-to-Device Communication Under Channel Uncertainties: A Stable Matching Approach}

\author{Monowar Hasan, {\em Student Member}, {\em IEEE}, and Ekram Hossain, {\em Fellow}, {\em IEEE}
\thanks{M. Hasan is with the Department of Computer Science, University of Illinois Urbana-Champaign, Illinois, USA; E. Hossain is with the Department of Electrical and Computer Engineering, University of Manitoba, Winnipeg, Canada (emails:  mhasan11@illinois.edu, Ekram.Hossain@umanitoba.ca).  This work was supported by the Natural Sciences and Engineering Research Council of Canada (NSERC) through a Strategic  Project Grant (STPGP 430285).}}

\maketitle

\begin{abstract}


Wireless device-to-device (D2D) communication underlaying cellular network is a promising concept  to improve user experience and resource utilization. Unlike traditional D2D communication where two mobile devices in the proximity establish a direct local link bypassing the base station, in this work we focus on  relay-aided D2D communication. Relay-aided transmission could enhance the performance of D2D communication when D2D user equipments (UEs) are  far apart  from each other and/or the quality of D2D link is not good enough for direct
communication. Considering the uncertainties in wireless links, we model and analyze the performance of a relay-aided D2D communication network, where the relay nodes serve both the cellular and D2D users. In particular, we formulate the radio resource allocation problem in a two-hop network to guarantee the data rate of the UEs while protecting other receiving nodes from interference. Utilizing time sharing strategy, we provide a centralized solution under bounded channel uncertainty. With a view to reducing the computational burden at relay nodes, we propose a distributed solution approach using stable matching to allocate radio resources in an efficient and computationally inexpensive way. Numerical results show that the performance of the proposed method is close to the centralized optimal solution and there is a distance margin beyond which relaying of D2D traffic improves network performance.

\end{abstract}


\section{Introduction}

We consider relay-assisted device-to-device (D2D) communication underlaying LTE-A cellular networks where D2D user equipments (UEs) are served by the relay nodes~\cite{d2d_our_paper,d2d_our_robust}. When the link condition between two D2D UEs is too poor for direct communication,  the D2D traffic can be transmitted via a relay node, which performs scheduling and resource allocation for the D2D UEs. We refer to this as \textit{relay-aided D2D communication} which can be an efficient approach to provide better quality of service (QoS) for communication between distant D2D UEs. For this, we utilize the LTE-A Layer-3 (L3) relay~ \cite{relay-book-1}. We consider scenarios in which the potential D2D UEs are located in the same macrocell; however, the proximity and link condition may not be favorable for direct communication. Therefore, they communicate via relays. The radio resources (e.g., resource blocks [RBs] and transmission power) at the relays are shared among the D2D communication links and the two-hop cellular links. 

We formulate an optimization problem to allocate radio resources at a relay node in a muti-relay multi-user orthogonal frequency-division multiple access (OFDMA) cellular network (e.g., LTE-A network). Due to the NP-hardness of the resource allocation problem, we utilize time sharing strategy and provide an asymptotically optimal centralized solution. Considering the random nature of wireless channels, we reformulate the resource allocation problem using the worst-case robust optimization theory. The uncertainties in link gains are modeled using ellipsoidal uncertainty sets. Each relay node can centrally solve the problem taking channel uncertainty into consideration. However, considering the high (e.g., cubic to the number of UEs and RBs) computational overhead at the relay nodes, we provide a \textit{distributed} solution based on \textit{stable matching} theory which is computationally inexpensive (e.g., linear with the number of UEs and RBs). We also analyze the stability, uniqueness, and optimality of the proposed solution. 

Considering the computational and signalling overheads and lack of scalability of the centralized solutions, game theoretical models have been widely used for wireless resource allocation problems. However, the analytical tractability of equilibrium in such game-theoretical models requires special properties for the objective functions, such as convexity, which may not be satisfied for many practical cases \cite{matching_walid_overview}. In this context, 
resource allocation  using matching theory has several beneficial properties \cite{sm_def,matching_walid_overview}. For example, the stable matching algorithm terminates for every given preference profile. The outcome of matching provides suitable solutions in terms of stability and optimality, which can accurately reflect different system objectives. Besides,  with suitable data structures, a Pareto optimal stable matching (e.g., allocation of resources to the UEs) can be obtained quickly for online implementation.

The goal of this work is to design a practical radio resource scheme for relay-aided D2D communication in a multi-user multi-relay OFDMA network. As opposed to most of the work in the literature where channel gain information is assumed to be perfect, we capture the dynamics of random and time-varying nature of wireless channels. To this end, we develop a low-complexity distributed solution based on the theory of stable matching and demonstrate how this scheme can be implemented in a practical LTE-A system. The major contributions of this work can be summarized as follows:

\begin{itemize}

\item We model and analyze the radio resource allocation problem for relay-aided D2D communication underlaying an OFDMA cellular network considering uncertainties in channel gains. We formulate an optimization problem to maximize system capacity in a two-hop network while satisfying the minimum data rate requirement for each UE and limiting the interference to  other receiving nodes. We show that the convexity of the optimization problem is conserved under bounded channel uncertainty in both the useful and interference links.

\item Using the theory of stable matching, we develop a distributed iterative solution, which considers bounded channel uncertainty. The stability, uniqueness, optimality, and complexity of the proposed solution are analyzed. We also present a possible implementation approach of our proposed scheme in an LTE-A system.

\item Numerical results show that the proposed distributed solution performs close to the upper bound of the optimal solution obtained in a centralized  manner; however, it incurs a lower (e.g., linear compared to cubic) computational complexity. Through simulations, we also compare the performance of the proposed approach with a traditional underlay D2D communication scheme and observe that after a distant margin, relaying of D2D traffic improves network performance.

\end{itemize}

We organize the rest of the paper as follows. We briefly review the related work in Section \ref{sec:related_works}.  Section \ref{sec:system_model} presents the system model and related assumptions. Followed by the formulation of the nominal resource allocation problem  in Section \ref{sec:rap_optimization}, we reformulate the resource allocation problem considering wireless channel gain uncertainties in Section \ref{sec:rap_uncert}. We develop the stable matching-based distributed resource allocation algorithm in Section \ref{sec:RA_SM}. 
Theoretical analysis of the proposed solution is presented in Section \ref{sec:analysis_sm}. In Section \ref{sec:numerical_results}, we present the performance evaluation results before we conclude the paper in Section \ref{sec:conclusion_sm}.

\section{Related Work} \label{sec:related_works}

\begin{table*}[!t]
\caption{Summary of Related Work and Comparison with Proposed Scheme}
\label{tab:summary}
\centering
\resizebox{1.0\textwidth}{!}{
\begin{tabular}{C{1.7cm}| P{2.8cm}| c| C{1.45cm}| P{3.6cm}| C{1.9cm}| P{1.8cm} }
\hline
\multicolumn{1}{c|}{\bfseries Reference} & \multicolumn{1}{c|}{\bfseries Problem focus} & \multicolumn{1}{c|}{\bfseries Relay-aided} & \bfseries Channel information & \multicolumn{1}{c|}{\bfseries Solution approach} & \multicolumn{1}{c|}{\bfseries Solution type} & \multicolumn{1}{c}{\bfseries Optimality}\\
\hline\hline
\parbox[t]{2mm}{\multirow{17}{*}{\hspace*{-2.5em}\rotatebox[origin=c]{90}{ \parbox{2.5cm}{\centering Work on D2D communication}}}} \hspace*{-1em}
\cite{zul-d2d} & Resource allocation & No & Perfect & Proposed greedy heuristic & Centralized & Suboptimal\\
\cite{phond-d2d} & Resource allocation & No & Perfect & Column generation based greedy heuristic & Centralized & Suboptimal \\
\cite{d2d_new_paper} & Resource allocation & No\textsuperscript{*} & Perfect & Numerical optimization & Semi-distributed & Pareto optimal \\
\cite{d2d_intf_graph} & Resource allocation & No & Perfect & Interference graph coloring & Centralized & Suboptimal \\
\cite{le_d2d} & Resource allocation & No & Perfect & Two-phase heuristic & Centralized & Suboptimal \\
\cite{d2d_swarm} & Resource allocation, mode selection & No & Perfect & Particle swarm optimization & Centralized & Suboptimal \\ 
\cite{d2d_relay_relselection} & Resource allocation, mode selection & Yes & Perfect & Proposed heuristic & Distributed & Suboptimal\\
\cite{d2d-rel-1} & Theoretical analysis, performance evaluation  & Yes & Perfect & Statistical analysis  & Centralized & Optimal\\
\cite{d2d_relay_2} & Performance evaluation  & Yes & Perfect & Heuristic, simulation  & Centralized & N/A\textsuperscript{$\ddagger$} \\
\cite{d2d_our_paper} & Resource allocation & Yes & Perfect & Numerical optimization & Centralized & Asymptotically optimal \\
\cite{d2d_our_robust} & Resource allocation & Yes & Uncertain\textsuperscript{\#} & Gradient-based iterative update & Distributed & Suboptimal \\
\hline
\parbox[t]{2mm}{\multirow{7}{*}{\hspace*{-2.5em}\rotatebox[origin=c]{90}{ \parbox{2.5cm}{\centering Work utilizes matching theory}}}} \hspace*{-1em}
\cite{macthinggame_cr}  & Resource allocation & N/A\textsuperscript{\textdagger} & Perfect & One-to-one matching & Centralized & Optimal \\
\cite{matching_queue}  & Cross layer scheduling & N/A\textsuperscript{\textdagger} & Perfect & Many-to-one matching & Centralized & N/A\textsuperscript{$\ddagger$} \\
\cite{matching_incmplt_tcom}  & Spectrum sharing & N/A\textsuperscript{\textdagger} & Complete, incomplete & One-to-one matching & Distributed & Pareto optimal \\
\cite{matching_walid_mehdi}   & Cell association & N/A\textsuperscript{\textdagger} & Perfect & Many-to-one matching & Distributed & N/A\textsuperscript{$\ddagger$} \\
\cite{matching_ul_dl}    & Resource allocation & N/A\textsuperscript{\textdagger} & Perfect & Many-to-one matching & Distributed & N/A\textsuperscript{$\ddagger$} \\
\cite{matching_fd}    & Resource allocation & N/A\textsuperscript{\textdagger} & Perfect & One-to-one matching & Centralized & N/A\textsuperscript{$\ddagger$} \\
\hline
\textit{Proposed scheme} & Resource allocation & Yes & Uncertain & Matching theory (many-to-one matching) & Distributed & Weak Pareto optimal \\
\hline
\multicolumn{6}{l}{\textsuperscript{*}\footnotesize{D2D UEs serve as relays to assist CUE-eNB communications.}} \\
\multicolumn{6}{l}{\textsuperscript{$\ddagger$}\footnotesize{No information is available.}} \\
\multicolumn{6}{l}{\textsuperscript{\#}\footnotesize{Uncertainty in channel gain in the direct link between UEs (relays) and relays (eNB) is not considered.}} \\
\multicolumn{6}{l}{\textsuperscript{\textdagger}\footnotesize{Not applicable for the considered system model.}} \\
\end{tabular} }
\end{table*}

Despite the fact that the resource allocation problems in cellular D2D communication have been intensively studied in the recent literature, only a very few work consider relays for D2D communication and incorporate wireless link uncertainties in the formulation of the resource allocation problem. In \cite{zul-d2d}, a greedy heuristic-based resource allocation scheme is proposed for both uplink and downlink scenarios where a D2D pair shares the same resources with cellular UE (CUE) only if the achieved signal-to-interference-plus-noise ratio (${\rm SINR}$) is greater than a given ${\rm SINR}$ requirement. A resource allocation scheme based on column generation method is proposed in \cite{phond-d2d} to maximize the spectrum utilization by finding the minimum transmission length (i.e., time slots) for D2D links while protecting the cellular users from interference and guaranteeing QoS. A new spectrum sharing protocol for D2D communication overlaying a cellular network is proposed in \cite{d2d_new_paper}, which allows the D2D users to communicate bi-directionally while assisting the two-way communications between the eNB and the CUE.  A graph-based resource allocation method for cellular networks with underlay D2D communication is proposed in \cite{d2d_intf_graph}.  A two-phase resource allocation scheme for cellular network with underlaying D2D communication is proposed in \cite{le_d2d}. In \cite{d2d_swarm}, the mode selection and resource allocation problem for D2D communication underlaying cellular networks is investigated and the solution is obtained by particle swarm optimization. The works above do not consider relays for D2D communication.

A distributed relay selection method for relay-assisted D2D communication system is proposed in \cite{d2d_relay_relselection}.  In \cite{d2d-rel-1,  d2d_relay_2}, authors investigate the maximum ergodic capacity and outage probability of cooperative relaying in relay-assisted D2D communication considering power constraints at the eNB.  Taking the advantage of L3 relays supported by the 3GPP standard, in  \cite{d2d_our_paper},  a centralized resource allocation approach is proposed for relay-assisted D2D communication assuming that perfect channel information is available. 
A gradient-based distributed resource allocation scheme is proposed in \cite{d2d_our_robust} for relay-aided D2D communication in  a multi-relay network under uncertain channel information. In this multi-relay network, the interference link gain between a UE and other relays (to which the UE is not associated with) is modeled with ellipsoidal uncertainty sets. However, the uncertainty in direct channel gain between relay and the UE is not considered. In this paper, we remodel the previous formulation and extend the work in \cite{d2d_our_robust} by incorporating uncertainties in \textit{both} the useful and interference links. In particular, we present a distributed resource allocation algorithm using stable matching considering the uncertainties in wireless channel gains (e.g., channel quality indicator [CQI] parameters according to the LTE-A terminology).

Although not in the context of D2D communication, matching theory has been used in the literature to address the radio resource allocation problems in wireless networks. A spectrum allocation algorithm using matching theory is proposed in \cite{macthinggame_cr} for a cognitive radio network (CRN) under perfect channel assumption. In \cite{matching_queue}, a two-sided stable matching algorithm is applied for adaptive multi-user scheduling in an LTE-A network. A distributed matching algorithm is proposed in \cite{matching_incmplt_tcom} for cooperative spectrum sharing among multiple primary and secondary users with incomplete information in a CRN. In \cite{matching_walid_mehdi}, a distributed algorithm is proposed to solve the user association problem in the downlink of small cell networks (SCNs). A matching-based subcarrier allocation approach is proposed in \cite{matching_ul_dl} for services with coupled uplink and downlink  QoS requirements. The radio resource (e.g., subcarrier and power) allocation problem for a full-duplex OFDMA network is modeled as a transmitter-receiver-subcarrier
matching problem in \cite{matching_fd}. 

The matching-based solutions proposed in \cite{macthinggame_cr, matching_queue, matching_incmplt_tcom, matching_walid_mehdi, matching_ul_dl, matching_fd} do not consider D2D-enabled networks. In the context of D2D communication, most of the works (e.g., \cite{zul-d2d, le_d2d, d2d_intf_graph, d2d_swarm, phond-d2d, d2d-rel-1, d2d_relay_2, d2d_our_paper}) provide centralized solutions. Also, note that in \cite{zul-d2d, d2d_new_paper, le_d2d, d2d_intf_graph, d2d_swarm, phond-d2d}, the effect of relaying on D2D communication is not investigated. Moreover, the wireless link uncertainty is not considered in \cite{zul-d2d, d2d_new_paper, le_d2d, d2d_intf_graph, d2d_swarm, phond-d2d, d2d_relay_relselection, d2d-rel-1,  d2d_relay_2, d2d_our_paper, macthinggame_cr, matching_queue, matching_walid_mehdi, matching_ul_dl, matching_fd}.  Different from the above works, we propose a stable matching-based distributed radio resource allocation approach considering the channel gain uncertainties in a multi-relay and multi-user relay-aided D2D communication scenario. A summary of the related work and comparison with our proposed approach is presented in Table \ref{tab:summary}.

\section{System Model} \label{sec:system_model}

\subsection{Network Model}

Let $\mathcal{L} = \lbrace 1, 2, \ldots, L \rbrace$ denote the set of fixed-location L3 relays in the network  (Fig.~1 in \cite{d2d_our_robust}). The system bandwidth is divided into $N$ orthogonal RBs denoted by $\mathcal{N} = \lbrace 1, 2, \ldots, N \rbrace$ which are used by all the relays in a spectrum underlay fashion. The set of CUEs and D2D pairs are denoted by $\mathcal{C} = \lbrace 1, 2, \ldots, C \rbrace$ and $\mathcal{D} = \lbrace 1, 2, \ldots, D \rbrace$, respectively. We assume that association of the UEs (both cellular and D2D)  to the corresponding relays are performed before resource allocation. Prior to resource allocation, D2D pairs are also discovered and the D2D session is setup by transmitting known synchronization or reference signals \cite{network_asst_d2d}. 

We assume that the CUEs are outside the coverage region of the eNB and/or having bad channel condition, and therefore, the CUE-eNB communications need to be supported by the relays.  Communication between two D2D UEs requires the assistance of a relay node due to poor propagation condition. The UEs assisted by relay $l$ are denoted by $u_l$. The set of UEs assisted by relay $l$ is $\mathcal{U}_l = \lbrace 1, 2,\ldots,  U_l \rbrace$ such that $\mathcal{U}_l \subseteq \lbrace \mathcal{C} \cup \mathcal{D} \rbrace, \forall l \in \mathcal{L}$, $\bigcup_l \mathcal{U}_l = \lbrace \mathcal{C} \cup \mathcal{D} \rbrace$, and  $\bigcap_l \mathcal{U}_l = \varnothing$.  

In the second hop, there could be multiple relays transmitting to their associated D2D UEs. We assume that multiple relays transmit to the eNB (in order to forward CUEs' traffic) using orthogonal channels and this scheduling of relays is done by the eNB\footnote{Scheduling of relay nodes by the eNB is not within the scope of this work.}.  In our system model, taking advantage of the capabilities of L3 relays, scheduling and resource allocation for the UEs is performed in the relay nodes to reduce the computational load at the eNB. 

\subsection{Achievable Data Rate}

Let $\gamma_{u_l, l, 1}^{(n)}$ denote the unit power ${\rm SINR}$ for the link between UE $u_l \in \mathcal{U}_l$ and relay $l$ using RB $n$ in the first hop and $\gamma_{l, u_l, 2}^{(n)}$ be the unit power ${\rm SINR}$ for the second hop. Note that, in the second hop, when the relays transmit CUEs' traffic  (i.e.,  $u_l \in \lbrace \mathcal{C} \cap \mathcal{U}_l \rbrace$),  $\gamma_{l, u_l, 2}^{(n)}$  denotes the  unit power ${\rm SINR}$ for the link between relay $l$ and the eNB. On the other hand, when a relay transmits to a D2D UE (i.e., $u_l \in \lbrace \mathcal{D} \cap \mathcal{U}_l \rbrace$),  $\gamma_{l, u_l, 2}^{(n)}$ refers to  the unit power ${\rm SINR}$ for the link between relay $l$ and the receiving D2D UE for the D2D-pair.

Let $P_{i, j}^{(n)} \geq 0$ denote the transmit power in the  link between $i$ and $j$ over RB $n$ and $B_{RB}$ is the bandwidth of an RB. The achievable data rate\footnote{We will present the rate expressions in Section \ref{sec:obj_function}.} for $u_l$ in the first hop can be expressed as $r_{u_l, 1}^{(n)} = B_{RB} \log_2 \left( 1 + P_{u_l, l}^{(n)} \gamma_{u_l, l, 1}^{(n)} \right)$. 
Similarly, the achievable data rate in the second hop is $r_{u_l, 2}^{(n)} = 
B_{RB} \log_2 \left( 1 + P_{l, u_l}^{(n)} \gamma_{l, u_l, 2}^{(n)} \right)$. Since we consider a two-hop communication, the end-to-end data rate\footnote{In a conventional D2D communication approach where two D2D UEs  communicate directly without a relay, the achievable data rate for D2D UE $u \in \mathcal{D}$ over RB $n$ can be expressed as $\widetilde{R}_{u}^{(n)} = B_{RB} \log_2 \left( 1 + P_{u}^{(n)} \tilde{\gamma}_{u}^{(n)} \right)$, where $\tilde{\gamma}_{u}^{(n)}  = \tfrac{h_{u, u}^{(n)}}{ \sum\limits_{\forall j \in \hat{{\mathcal U}}_u} P_{j}^{(n)} g_{u, j}^{(n)} + \sigma^2} $, $h_{u, u}^{(n)}$ is the channel gain in the link between the D2D UEs and  $\hat{{\mathcal U}}_u$ denotes the set of UEs transmitting using the same RB(s) as $u$. \label{ftn:direct_d2d_rate}} for  $u_l$ on RB $n$ is half of the minimum achievable data rate over two hops \cite{mutihop-rate}, i.e.,   
\begin{equation}
\label{eqn:e2e_rate}
R_{u_l}^{(n)} =  \frac{1}{2} \min \left\lbrace  r_{u_l, 1}^{(n)} , r_{u_l, 2}^{(n)}    \right\rbrace.
\end{equation}

\section{Resource Allocation: Formulation of the Nominal Problem} \label{sec:rap_optimization}

In the following, we present the formulation of the resource allocation problem assuming that perfect channel gain information is available. This formulation is referred to as the nominal problem since the uncertainties in channel gains are not considered.

For each relay, the objective of radio resource (i.e., RB and transmit power) allocation is to obtain  the assignment of RB and power level to the UEs that maximizes the system capacity, which is defined as the minimum achievable data rate over two hops. Let the maximum allowable transmit power for UE (relay) is $P_{u_l}^{max}$ ($P_l^{max}$) and let the QoS (i.e., data rate) requirement for UE $u_l$ be denoted by $Q_{u_l}$. The RB allocation indicator is a binary decision variable $x_{u_l}^{(n)} \in \lbrace 0, 1\rbrace$, where 
\begin{equation} \label{eq:bin_con}
x_{u_l}^{(n)}  = \begin{cases}
 1, \quad  \text{if RB $n$ is assigned to UE $u_l$} \\
 0, \quad \text{otherwise.}
\end{cases}
\end{equation}

\subsection{Objective Function} \label{sec:obj_function}

Let $\displaystyle R_{u_l} = \sum_{n =1}^N  x_{u_l}^{(n)} R_{u_l}^{(n)}$ denote the achievable sum-rate over allocated RB(s).  We consider that the same RB(s) will be used by the relay in both the hops (i.e., for communication between relay and eNB and between relay and D2D UEs). The objective of resource allocation problem is to maximize the end-to-end rate for each relay $l \in \mathcal{L}$  as follows:
\begin{equation}
\underset{x_{u_l}^{(n)}, P_{u_l, l}^{(n)}, P_{l, u_l}^{(n)}}{\operatorname{max}} ~ \sum_{u_l \in \mathcal{U}_l }  \sum_{n =1}^N   x_{u_l}^{(n)} R_{u_l}^{(n)} \label{eq:obj_func}
\end{equation}
where the rate of UE $u_l$ over RB $n$ 
$$R_{u_l}^{(n)} = \frac{1}{2} \min \left\lbrace \begin{matrix}
B_{RB} \log_2 \left( 1 + P_{u_l, l}^{(n)} \gamma_{u_l, l, 1}^{(n)} \right)  \vspace*{0.5em}\\
B_{RB} \log_2 \left( 1 + P_{l, u_l}^{(n)} \gamma_{l, u_l, 2}^{(n)} \right)   
\end{matrix} \right\rbrace.$$

In (\ref{eq:obj_func}), the unit power ${\rm SINR}$  for the first hop, 
\begin{equation} \label{eq:SINR1_w_X}
\gamma_{u_l, l, 1}^{(n)} = \frac{h_{u_l, l, 1}^{(n)}}{  \sum\limits_{\substack{\forall u_j \in \mathcal U_j, \\ j \neq l, j \in \mathcal{L}}}  x_{u_j}^{(n)} P_{u_j, j}^{(n)} g_{u_j, l, 1}^{(n)} + \sigma^2  }
\vspace{-0.5em}
\end{equation}
where $h_{i,j, k}^{(n)}$ denotes the direct link gain between node $i$ and $j$ over RB $n$ for hop $k \in \lbrace 1,2 \rbrace$, $\sigma^2 =  N_0 B_{RB}$ in which $N_0$ denotes thermal noise. The interference link gain between relay (UE) $i$  and UE (relay) $j$ over RB $n$ in hop $k$ is denoted by $g_{i,j, k}^{(n)}$,  where UE (relay) $j$ is not associated with relay (UE) $i$.  Similarly, the unit power ${\rm SINR}$ for the second hop\footnote{According to LTE-A standard, the L3 relays are able to peform similar operation as an eNB. Besides, the relays in the network are interconnected through X2 interface for better interference management \cite{lte_arch}. Since the relays can estimate the CQI values (and hence the interference level) using X2 interface, it is straightforward to account for interference in  (\ref{eq:SINR1_w_X}) and (\ref{eq:SINR2_w_X}). Consequently, interference from other transmitter nodes (e.g., UEs associated to other relays in the first hop or other relays in the second hop) will appear as a constant term in (\ref{eq:SINR1_w_X}) and (\ref{eq:SINR2_w_X}). \label{ftnt:X2_constant}}, \vspace{-0.3em} 
\begin{equation} \label{eq:SINR2_w_X}
\gamma_{l, u_l, 2}^{(n)} = \begin{cases}
\frac{h_{l, u_l, 2}^{(n)}}{ \sum\limits_{\substack{\forall u_j \in \lbrace \mathcal{D} \cap \mathcal{U}_j \rbrace, \\ j \neq l, j \in \mathcal{L}}} \hspace{-0.5em} x_{u_j}^{(n)} P_{j, u_j}^{(n)} g_{j, eNB, 2}^{(n)} + \sigma^2}, ~  u_l \in \lbrace \mathcal{C} \cap \mathcal{U}_l \rbrace \\
\frac{h_{l, u_l, 2}^{(n)}}{\sum\limits_{\substack{\forall u_j \in  \mathcal{U}_j , \\ j \neq l, j \in \mathcal{L}}} x_{u_j}^{(n)}  P_{j, u_j}^{(n)} g_{j, u_l, 2}^{(n)} + \sigma^2}, \quad ~~~~ u_l \in \lbrace \mathcal{D} \cap \mathcal{U}_l \rbrace
\end{cases}
\end{equation}
where $h_{l, u_l, 2}$ denotes the channel gain between relay-eNB link for CUEs (e.g., $u_l \in \lbrace \mathcal{C} \cap \mathcal{U}_l \rbrace$) or the channel gain between relay and receiving D2D UEs (e.g., $u_l \in \lbrace \mathcal{D} \cap \mathcal{U}_l \rbrace$).
From (\ref{eqn:e2e_rate}), the maximum data rate for UE $u_l$ over RB $n$ is achieved when $P_{u_l, l}^{(n)} \gamma_{u_l, l, 1}^{(n)} = P_{l, u_l}^{(n)} \gamma_{l, u_l, 2}^{(n)}$. Therefore, in the second hop, the power $P_{l, u_l}$ allocated for UE $u_l$, can be expressed as a function of power allocated for transmission in the first hop, $P_{u_l, l}$ as follows: $P_{l, u_l}^{(n)}  = \frac{\gamma_{u_l, l, 1}^{(n)}}{\gamma_{l, u_l, 2}^{(n)}}P_{u_l, l}^{(n)} \approx \frac{h_{u_l, l, 1}^{(n)}}{h_{l, u_l, 2}^{(n)}}P_{u_l, l}^{(n)} $. Hence, the data rate for $u_l$ over RB $n$ can  be expressed as 
$R_{u_l}^{(n)} = \frac{1}{2} B_{RB}  \log_2 \left( 1 + P_{u_l, l}^{(n)} \gamma_{u_l, l, 1}^{(n)} \right)$. Considering the above, the objective function in (\ref{eq:obj_func}) can be rewritten as
\begin{equation}
\underset{x_{u_l}^{(n)}, P_{u_l, l}^{(n)}}{\operatorname{max}} ~ \sum_{u_l \in \mathcal{U}_l } \sum_{n =1}^N   \tfrac{1}{2}  x_{u_l}^{(n)}   B_{RB} \log_2   \left(  1 +   P_{u_l, l}^{(n)} \gamma_{u_l, l, 1}^{(n)}   \right)  \label{eq:obj_func-relx}.
\end{equation}

For each relay $l \in \mathcal{L}$ in the network, the objective of resource allocation problem is to obtain the RB and power allocation vectors, i.e., 
$\mathbf{x}_{\boldsymbol l} = \left[ x_{1}^{(1)}, \ldots,  x_{1}^{(N)}, \ldots, x_{U_l}^{(1)}, \ldots, x_{U_l}^{(N)} \right]^\mathsf{T}$ and $\mathbf{P}_{\boldsymbol l} = \left[ P_{1,l}^{(1)}, \ldots, P_{1,l}^{(N)}, \ldots,  P_{U_l,l}^{(1)}, \ldots, P_{U_l,l}^{(N)} \right]^\mathsf{T}$, respectively, which maximize the data rate. 

\subsection{Constraint Sets} \label{sec:const}

In order to ensure the required data rate for the UEs while protecting all receiver nodes from harmful interference, we define the following set of constraints.

\begin{itemize}

\item The constraint in (\ref{eq:con-bin}) ensures that each RB is assigned to only one UE, i.e.,
\setlength{\arraycolsep}{0.11em}
\begin{eqnarray}
\displaystyle \sum_{u_l \in \mathcal{U}_l} x_{u_l}^{(n)} &\leq&  1, ~ \forall n \in \mathcal{N}. \label{eq:con-bin} 
\end{eqnarray}
\setlength{\arraycolsep}{5pt}

\item The following constraints limit the transmit power in each of the hops to the maximum power budget:
\begin{align}
\sum_{n =1}^N x_{u_l}^{(n)} P_{u_l, l}^{(n)}  &\leq P_{u_l}^{max}, ~ \forall u_l \in \mathcal{U}_l \label{eq:con-pow-ue} \\
\sum_{u_l \in \mathcal{U}_l } \sum_{n =1}^N x_{u_l}^{(n)} P_{l, u_l}^{(n)}  &\leq ~ P_l^{max}. \label{eq:con-pow-rel}
\end{align}

\item Similar to \cite{amr_icc}, we assume that there is a maximum tolerable interference threshold limit for each allocated RB. The constraints in (\ref{eq:con-intf-1}) and (\ref{eq:con-intf-2}) limit the amount of interference introduced to the other relays and the receiving D2D UEs in the first and second hop, respectively, to be less than some threshold, i.e., 
\begin{align}
\sum_{u_l \in \mathcal{U}_l } x_{u_l}^{(n)} P_{u_l, l}^{(n)} g_{{u_l^*}, l, 1}^{(n)} &\leq I_{th, 1}^{(n)},  ~~\forall n \in \mathcal{N} \label{eq:con-intf-1}\\
\sum_{u_l \in \mathcal{U}_l } x_{u_l}^{(n)}  P_{l, u_l}^{(n)} g_{l, {u_l^*}, 2}^{(n)} &\leq  I_{th, 2}^{(n)}, ~~\forall n \in \mathcal{N} \label{eq:con-intf-2}.
\end{align}

\item The minimum data rate requirements for the CUE and D2D UEs is ensured by the following constraint: 
\begin{equation}
\quad R_{u_l} \geq Q_{u_l}, ~ \forall u_l \in \mathcal{U}_l \label{eq:con-QoS-cue}.
\end{equation}

\item The binary  decision variable on RB allocation and non-negativity condition of transmission power is defined by
\begin{eqnarray}
x_{u_l}^{(n)} \in \lbrace 0, 1 \rbrace, ~~ P_{u_l, l}^{(n)} \geq 0, ~~~\forall u_l \in \mathcal{U}_l, \forall n \in \mathcal{N}. \label{eq:x-s}
\end{eqnarray}

\end{itemize}

Note that in constraint (\ref{eq:con-intf-1}) and (\ref{eq:con-intf-2}), we adopt the concept of reference user. For example, to allocate the power level considering the interference threshold in the first hop, each UE $u_l$ associated with relay node $l$  obtains the reference user $u_l^*$ associated with the other relays and the corresponding channel gain $g_{{u_l^*}, l, 1}^{(n)}$ for $\forall n$  according to the following equation:
\begin{equation}
u_{l}^* = \underset{j}{\operatorname{argmax}} ~ g_{u_l , j, 1}^{(n)},~~ u_l \in \mathcal{U}_l, j \neq l, j \in \mathcal{L} \label{eq:ref_user1}.
\end{equation}
Similarly, in the second hop, for each relay $l$, the transmit power will be adjusted accordingly considering interference introduced to the receiving D2D UEs (associated with other relays) considering the corresponding channel gain $g_{l, {u_l^*}, 2}^{(n)}$ for $\forall n$, where the reference user  is obtained by 
\begin{equation}
u_{l}^* = \underset{u_j}{\operatorname{argmax}} ~ g_{l , u_j, 2}^{(n)}, ~~ j \neq l, j \in \mathcal{L},  u_j \in \lbrace \mathcal{D} \cap \mathcal{U}_j \rbrace. \label{eq:ref_user2}
\end{equation}   

\subsection{Centralized Solution}

\begin{corollary}
\label{cor:MINLP}
The objective function in (\ref{eq:obj_func-relx}) and the set of constraints in (\ref{eq:con-bin})-(\ref{eq:x-s}) turn the optimization problem to a mixed-integer non-linear program (MINLP) with non-convex feasible set. Therefore, the formulation described in Section \ref{sec:rap_optimization} is NP-hard. 
\end{corollary} 

A well-known approach to solve the above problem is to relax the constraint that an RB is used by only one UE by using the time-sharing factor \cite{relax-con-1}. In particular, we relax the optimization problem by replacing the non-convex constraint $x_{u_l}^{(n)} \in \lbrace 0,  1\rbrace$ with the convex constraint $0 < x_{u_l}^{(n)}  \leq 1$. Thus $x_{u_l}^{(n)}$ represents the sharing factor where each $x_{u_l}^{(n)}$ denotes the portion of time that RB $n$ is assigned to UE $u_l$ and satisfies the constraint $\displaystyle \sum_{u_l \in \mathcal{U}_l} x_{u_l}^{(n)} \leq 1, ~\forall n$. Besides, we introduce a new variable $S_{u_l, l}^{(n)} = x_{u_l}^{(n)} P_{u_l,l}^{(n)} \geq 0$, which denotes the actual transmit power of UE $u_l$ on RB $n$ \cite{time-share-1}. Then the relaxed problem can be stated as follows:  
\setlength{\arraycolsep}{0.0em}
\begin{eqnarray}
\mathbf{(P2)}  \hspace*{2.5em}  
\underset{x_{u_l}^{(n)}, S_{u_l, l}^{(n)}}{\operatorname{max}} ~ \sum_{u_l \in \mathcal{U}_l }   R_{u_l} \hspace*{-1.9em} \label{eq:relx_objfunc} \\ 
\text{subject to} \quad \text{(\ref{eq:con-bin}), (\ref{eq:con-QoS-cue})}  \text{~~and~} \hspace{-4.0em}\nonumber \\
\sum_{n =1}^N S_{u_l, l}^{(n)} &~~{\leq}~ P_{u_l}^{max}, ~\forall u_l  \hspace*{2.0em} \label{eq:con-pow-ue-relx} \\
 \sum_{u_l \in \mathcal{U}_l } \sum_{n =1}^N H_{u_l, l}^{(n)} S_{u_l, l}^{(n)} &~{\leq}~ P_l^{max}~~~~~  \hspace*{2.0em} \label{eq:con-pow-rel-relx} \\
 \sum_{u_l \in \mathcal{U}_l } S_{u_l, l}^{(n)} g_{{u_l^*}, l, 1}^{(n)} &~{\leq}~ I_{th, 1}^{(n)}, ~~ \forall n \hspace*{2em}\label{eq:con-intf-1-relx}\\
 \sum_{u_l \in \mathcal{U}_l } H_{u_l, l}^{(n)} S_{u_l, l}^{(n)} g_{l, {u_l^*}, 2}^{(n)} &~{\leq}~ I_{th, 2}^{(n)}, ~~\forall n \hspace*{2em} \label{eq:con-intf-2-relx} \\
 0 ~<~ x_{u_l}^{(n)} ~\leq~ 1, ~S_{u_l, l}^{(n)} \hspace{0em} &{\geq} ~ 0,  ~~~~~~\forall n, u_l \hspace*{0.5em} \label{eq:x-s-relx}
\end{eqnarray}
\setlength{\arraycolsep}{5pt}
where $\gamma_{u_l, l, 1}^{(n)} = \tfrac{h_{u_l, l}^{(n)}}{ \sum\limits_{\substack{\forall u_j \in \mathcal U_j, \\ j \neq l, j \in \mathcal{L}}}   S_{u_j, j}^{(n)} g_{u_j, l, 1}^{(n)} + \sigma^2  }$, $H_{u_l, l}^{(n)} = \tfrac{h_{u_l, l, 1}^{(n)}}{h_{l, u_l, 2}^{(n)}}$ and $ R_{u_l} = \displaystyle \sum_{n =1}^N  \tfrac{1}{2} x_{u_l}^{(n)} B_{RB} \log_2  \left(  1 +   \tfrac{S_{u_l, l}^{(n)} \gamma_{u_l, l, 1}^{(n)}} {x_{u_l}^{(n)} } \right)$.  

\begin{corollary}
\label{cor:relx-convex}
The objective function in (\ref{eq:relx_objfunc}) is concave, the constraint in (\ref{eq:con-QoS-cue}) is convex, and the remaining constraints in (\ref{eq:con-bin}), (\ref{eq:con-pow-ue-relx})-(\ref{eq:x-s-relx}) are affine. Therefore, the optimization problem $\mathbf{P2}$ is convex.
\end{corollary}

Since $\mathbf{P2}$ is a non-linear convex problem, each relay can solve the optimization problem using standard algorithms such as interior point method \cite[Chapter 11]{book-boyd}. The centralized optimization-based solution is summarized in \textbf{Algorithm \ref{alg:cent_soln}}. Each relay locally solves the optimization problem $\mathbf{P2}$ and informs the other relays the allocation vectors using X2 interface. The process is repeated until the data rate is maximized, e.g., $R_l(t) - R_l(t-1) < \epsilon$ for $\forall l$, where $R_l(\cdot) = \displaystyle \sum_{u_l \in \mathcal{U}_l} R_{u_l}(\cdot)$ is the sum data rate for relay $l$ obtained by solving the optimization problem at iteration $(\cdot)$ and $\epsilon$ is a small value. 

The duality gap of any optimization problem satisfying the time-sharing condition becomes negligible as the number of RBs becomes significantly large. The optimization problem $\mathbf{P2}$ satisfies the time-sharing condition, and therefore, the solution of the relaxed problem is asymptotically optimal \cite{large-rb-dual}.

Given the parameters of other relays (e.g., $\mathbf{x}_{\boldsymbol j}$,  $\mathbf{P}_{\boldsymbol j} ~\forall j \neq l, j \in \mathcal{L}$), at each iteration of \textbf{Algorithm \ref{alg:cent_soln}} the allocation vectors (e.g., $\mathbf{x}_{\boldsymbol l}$,  $\mathbf{P}_{\boldsymbol l}$) obtained at each relay $l$ provide locally optimal solution for $l$. In addition, \textbf{Algorithm \ref{alg:cent_soln}} allows the relays to perform allocation repeatedly with a view to finding the best possible allocation. If the data rate at the $(t+1)$-th iteration is not improved compared to that in the previous iteration $t$, the algorithm terminates, and the allocation at iteration $t$ will be the resultant solution. 
Each of the iteration of \textbf{Algorithm \ref{alg:cent_soln}}  outputs the solution of relaxed version of the original NP-hard optimization problem. Since the solution of relaxed problem gives us the upper bound, and at the termination of \textbf{Algorithm \ref{alg:cent_soln}}, we obtain an upper bound of the achievable sum rate.

\begin{algorithm} [!t]
\AtBeginEnvironment{algorithmic}{\small} 
\caption{Optimization-based resource allocation}
\label{alg:cent_soln}
\begin{algorithmic}[1]   
\renewcommand{\algorithmicrequire}{\textbf{Input:}}
\renewcommand{\algorithmicensure}{\textbf{Output:}}
\renewcommand{\algorithmicforall}{\textbf{for each}}
\renewcommand{\algorithmiccomment}[1]{\textit{/* #1 */}}

\STATE  Each relay  $l \in \mathcal{L}$ estimates the CQI values from previous time slot and determines reference gains $g_{u_l^*, l, 1}^{(n)}$ and  $g_{l, u_l^*, 2}^{(n)}, \forall u_l, n$.

\STATE Initialize $t:=0$.

\REPEAT 

\STATE Update $t:= t + 1$.

\STATE Each relay $l \in \mathcal{L}$:

\begin{itemize}
\setlength{\itemsep}{0pt}
\item solves the optimization problem $\mathbf{P2}$ and multicasts the allocation variables $\mathbf{x}_{\boldsymbol l}$,  $\mathbf{P}_{\boldsymbol l}$ to all relay $j \neq l, j \in \mathcal{L}$ over X2 interface. 

\item calculates the achievable data rate based on current allocation as  $\displaystyle R_l(t) := \sum_{u_l \in \mathcal{U}_l} R_{u_l}(t)$.
\end{itemize}

\UNTIL data rate not maximized \AND $t < T_{max}$.

\STATE  Allocate resources (i.e., RB and transmit power) to associated UEs for each relay.

\end{algorithmic}
\end{algorithm}

\section{Resource Allocation Under Channel Uncertainty} \label{sec:rap_uncert}

For worst-case robust resource optimization problems, the channel gain is assumed to have a bounded uncertainty of unknown distribution. An ellipsoid is often used  (e.g.,\cite{ellipsoid_ex1, ellipsoid_ex2, ellipsoid_ex3}) to approximate such an uncertainty region. 

\subsection{Uncertainty Sets} \label{subsec:uncert_set}

Let the variable $F_{u_l, u_j, l}^{(n)}$ denote the normalized channel gain which is defined as follows:
\begin{equation} \label{eq:ch_gain_f}
F_{u_l, u_j, l}^{(n)}  \triangleq  \frac{g_{u_j, l,1}^{(n)}}{h_{u_l, l, 1}^{(n)}}, \quad  \forall u_j \in \mathcal U_j, j \neq l, j \in \mathcal{L}.
\end{equation}
In addition, let $\mathcal{F}_{u_l, l}^{(n)}$ denote the uncertainty set that describes the perturbation of link gains for $u_l$ over RB $n$. The normalized gain is then denoted by 
\begin{equation} \label{eq:F_uncert}
F_{u_l, u_j, l}^{(n)} = \bar{F}_{u_l, u_j, l}^{(n)} + \Delta F_{u_l, u_j, l}^{(n)}
\end{equation}
where $\bar{F}_{u_l, u_j, l}^{(n)} $ is the nominal value and $\Delta F_{u_l, u_j, l}^{(n)}$ is the perturbation part. The uncertainty in the CQI values  is modeled under an ellipsoidal approximation as follows:
\begin{equation} \label{eq:uncert_set_f}
\hspace{-0.2em}
\mathcal{F}_{u_l, l}^{(n)} = \hspace{-0.2em} \left\lbrace \bar{F}_{u_l, u_j, l}^{(n)} + \Delta F_{u_l, u_j, l}^{(n)} : \hspace*{-0.8em} \sum_{\substack{\forall u_j \in \mathcal{U}_j, \\ j \neq l, j \in \mathcal{L} }} \hspace*{-0.7em} |\Delta F_{u_l, u_j, l}^{(n)}|^2 \leq \xi_{1_{u_l}}^{(n)}; \forall u_l, n \right\rbrace
\end{equation}
where $\xi_{1_{u_l}}^{(n)} \geq 0$ is the uncertainty bound in each RB. Using (\ref{eq:ch_gain_f}), we rewrite the rate expression for $u_l$ over RB $n$ as
\begin{equation} \label{eq:F_rate}
R_{u_l}^{(n)} = \tfrac{1}{2} B_{RB}  \log_2 \Bigg(  1 + \tfrac{P_{u_l, l}^{(n)}}{\sum\limits_{\substack{\forall u_j \in \mathcal{U}_j, \\ j \neq l, j \in \mathcal{L} }} \hspace{-0.4em} F_{u_l, u_j, l}^{(n)} P_{u_j, j}^{(n)} + \tilde{\sigma}_{u_l}^{(n)}} \Bigg)
\end{equation}
where $\tilde{\sigma}_{u_l}^{(n)} \triangleq \frac{\sigma^2}{h_{u_l, l, 1}^{(n)}}$ and $F_{u_l, u_j, l}^{(n)}$ is given by (\ref{eq:F_uncert}).

\subsection{Reformulation of the Optimization Problem Considering Channel Uncertainty}

Utilizing  uncertainty sets similar to (\ref{eq:uncert_set_f}) in the  constraints (\ref{eq:con-pow-rel-relx})-(\ref{eq:con-intf-2-relx}), the optimization problem $\mathbf{P2} $  can be equivalently represented under channel uncertainty as follows:
\setlength{\arraycolsep}{0.0em}
\begin{eqnarray}
\mathbf{(P3)} \hspace{4.0em} \underset{x_{u_l}^{(n)},S_{u_l, l}^{(n)}}{\operatorname{max}} \underset{\substack{\Delta F_{u_l, u_j, l}^{(n)}, \Delta g_{{u_l^*}, l, 1}^{(n)}, \\ \Delta H_{u_l, l}^{(n)}, \Delta H_{u_l, l}^{(n)}  g_{l, {u_l^*}, 2}^{(n)}}} {\operatorname{min}}  \sum_{u_l \in \mathcal{U}_l }  R_{u_l} \hspace{-6.5em} \label{eq:relx_objfunc_uncrt} \\ 
\text{subject to} \quad \text{(\ref{eq:con-bin}), (\ref{eq:con-QoS-cue}), (\ref{eq:con-pow-ue-relx}), (\ref{eq:x-s-relx})}  \text{~~and~} \hspace{-5.0em}\nonumber \\
 \sum_{u_l \in \mathcal{U}_l } \sum_{n =1}^N \left( \bar{H}_{u_l, l}^{(n)}  + \Delta H_{u_l, l}^{(n)}  \right) S_{u_l, l}^{(n)} &~{\leq}~ P_l^{max}~~~~~  \hspace*{2.0em} \label{eq:con-pow-rel-relx_uncrt} \\
 \sum_{u_l \in \mathcal{U}_l } \left( \bar{g}_{{u_l^*}, l, 1}^{(n)} + \Delta g_{{u_l^*}, l, 1}^{(n)} \right) S_{u_l, l}^{(n)}  &~{\leq}~ I_{th, 1}^{(n)}, ~~ \forall n \hspace*{2em}\label{eq:con-intf-1-relx_uncrt}\\
 \sum_{u_l \in \mathcal{U}_l } \left( \bar{H}_{u_l, l}^{(n)}  \bar{g}_{l, {u_l^*}, 2}^{(n)} + \Delta H_{u_l, l}^{(n)}  g_{l, {u_l^*}, 2}^{(n)} \right) S_{u_l, l}^{(n)} &~{\leq}~ I_{th, 2}^{(n)}, ~~\forall n \hspace*{2em} \\ \label{eq:con-intf-2-relx_uncrt}
\hspace{-0.0em}  \sum_{\substack{\forall u_j \in \mathcal{U}_j,  j \neq l, j \in \mathcal{L} }} \hspace{-0.5em} |\Delta F_{u_l, u_j, l}^{(n)} |^2 & ~{\leq} \big(\xi_{1_{u_l}}^{(n)}\big)^2\!, \!\forall u_l, n~ \hspace*{1.5em} \label{eq:bound_relx_objfunc_uncrt} \\
  \sum_{u_l \in \mathcal{U}_l } \sum_{n =1}^N |\Delta H_{u_l, l}^{(n)}|^2 &~{\leq}~ ({\xi_2}_{l})^2 ~~~~~~~~~ \hspace*{1.0em} \label{eq:bound_con-pow-rel-relx_uncrt} \\
 \sum_{u_l \in \mathcal{U}_l }  |\Delta g_{{u_l^*}, l, 1}^{(n)} | ^2&~{\leq}~ \big(\xi_{3_{u_l}}^{(n)}\big)^2, ~\forall n \hspace*{1.7em}\label{eq:bound_con-intf-1-relx_uncrt}\\  
 \sum_{u_l \in \mathcal{U}_l } |\Delta H_{u_l, l}^{(n)}  g_{l, {u_l^*}, 2}^{(n)}| ^2  &~{\leq}~ \big(\xi_{4_{u_l}}^{(n)}\big)^2, ~\forall n \hspace*{1.7em}  \label{eq:bound_con-intf-2-relx_uncrt}
\end{eqnarray}
\setlength{\arraycolsep}{5pt}
where for any parameter $y$, $\bar{y}$ denotes the nominal value and $\Delta y$ represents the corresponding deviation part; ${\xi_2}_{l},~ \xi_{3_{u_l}}^{(n)}$, and~$\xi_{4_{u_l}}^{(n)}$ are the maximum deviations (e.g., uncertainty bounds) of corresponding entries in CQI values. In $\mathbf{P3}$, $R_{u_l}$ is given by
\begin{eqnarray} \label{eq:rate_uncert_relx}
R_{u_l} = \sum\limits_{n =1}^N  \tfrac{1}{2} x_{u_l}^{(n)} B_{RB} \times \hspace{12em} \nonumber \\
\log_2 \Bigg(  1 + \tfrac{\tfrac{S_{u_l, l}^{(n)}}{x_{u_l}^{(n)} }}{\sum\limits_{\substack{\forall u_j \in \mathcal{U}_j, \\ j \neq l, j \in \mathcal{L} }} \hspace{-0.4em} \left( \bar{F}_{u_l, u_j, l}^{(n)} + \Delta F_{u_l, u_j, l}^{(n)}\right) S_{u_j, j}^{(n)} + \tilde{\sigma}_{u_l}^{(n)}} \Bigg). \hspace*{-0.6em}
\end{eqnarray}

The above optimization problem is subject to an infinite number of constraints with respect to the uncertainty sets and hence becomes a 
semi-infinite programming (SIP) problem \cite{sip}. In order to solve the SIP problem it is required to transform $\mathbf{P3}$ into an equivalent problem with finite number of constraints. Similar to \cite{ellipsoid_ex1, ellipsoid_ex2}, we apply the  Cauchy-Schwarz inequality \cite{cauchy} and transform the SIP problem. More specifically, utilizing Cauchy-Schwarz inequality,  we obtain the following:
\setlength{\arraycolsep}{1pt}
\begin{eqnarray} \label{eq:worst_bound_rate}
\sum_{\substack{\forall u_j \in \mathcal{U}_j, \\ j \neq l, j \in \mathcal{L} }} \hspace{-0.4em} \Delta F_{u_l, u_j, l}^{(n)} S_{u_j, j}^{(n)} 
&\leq& \sqrt{\sum_{\substack{\forall u_j \in \mathcal{U}_j, \\ j \neq l, j \in \mathcal{L} }} \hspace{-0.4em} |\Delta F_{u_l, u_j, l}^{(n)} |^2 \sum_{\substack{\forall u_j \in \mathcal{U}_j, \\ j \neq l, j \in \mathcal{L} }} \hspace{-0.4em} |S_{u_j, j}^{(n)} |^2 } \nonumber \\
&\leq & \xi_{1_{u_l}}^{(n)} \sqrt{\sum_{\substack{\forall u_j \in \mathcal{U}_j, \\ j \neq l, j \in \mathcal{L} }} \left( S_{u_j, j}^{(n)} \right)^2}. 
\end{eqnarray} 
\setlength{\arraycolsep}{5pt}
Similarly,
\begin{align}
\sum_{u_l \in \mathcal{U}_l } \sum_{n =1}^N \Delta H_{u_l, l}^{(n)} S_{u_l, l}^{(n)} &\leq  {\xi_2}_{l} \sqrt{ \sum_{u_l \in \mathcal{U}_l } \sum_{n =1}^N\left( S_{u_l, l}^{(n)}  \right)^2} \\
 \sum_{u_l \in \mathcal{U}_l } \Delta g_{{u_l^*}, l, 1}^{(n)}  S_{u_l, l}^{(n)}  & \leq \xi_{3_{u_l}}^{(n)} \sqrt{  \sum_{u_l \in \mathcal{U}_l }  \left( S_{u_l, l}^{(n)}  \right)^2} \\
 \sum_{u_l \in \mathcal{U}_l } \Delta H_{u_l, l}^{(n)}  g_{l, {u_l^*}, 2}^{(n)}  S_{u_l, l}^{(n)} & \leq \xi_{4_{u_l}}^{(n)} \sqrt{  \sum_{u_l \in \mathcal{U}_l } \left( S_{u_l, l}^{(n)}  \right)^2}. \label{eq:worst_bound_intf2}
\end{align}

\begin{figure*}[!t]
\normalsize

\setlength{\arraycolsep}{0.0em}
\begin{eqnarray}
\mathbf{(P4)} \hspace{4.0em} \underset{x_{u_l}^{(n)},S_{u_l, l}^{(n)}}{\operatorname{max}} \sum_{u_l \in \mathcal{U}_l }  R_{u_l} \hspace{5em}
 \label{eq:relx_objfunc_uncrt_CS} \\ 
\text{subject to} \quad \text{(\ref{eq:con-bin}), (\ref{eq:con-QoS-cue}), (\ref{eq:con-pow-ue-relx}), (\ref{eq:x-s-relx})}  \text{~~and~} \hspace{-3.0em}\nonumber \\
 \sum_{u_l \in \mathcal{U}_l } \sum_{n =1}^N \bar{H}_{u_l, l}^{(n)} S_{u_l, l}^{(n)} + {\xi_2}_{l} \sqrt{ \sum_{u_l \in \mathcal{U}_l } \sum_{n =1}^N\left( S_{u_l, l}^{(n)}  \right)^2}   &~{\leq}~ P_l^{max}~~~~~  \hspace*{2.0em} \label{eq:con-pow-rel-relx_uncrt-CS} \\
 \sum_{u_l \in \mathcal{U}_l } \bar{g}_{{u_l^*}, l, 1}^{(n)} S_{u_l, l}^{(n)} + \xi_{3_{u_l}}^{(n)} \sqrt{  \sum_{u_l \in \mathcal{U}_l }  \left( S_{u_l, l}^{(n)}  \right)^2} &~{\leq}~ I_{th, 1}^{(n)}, ~~ \forall n \hspace*{2em}\label{eq:con-intf-1-relx_uncrt-CS}\\
 \sum_{u_l \in \mathcal{U}_l }  \bar{H}_{u_l, l}^{(n)}  \bar{g}_{l, {u_l^*}, 2}^{(n)} S_{u_l, l}^{(n)} +  \xi_{4_{u_l}}^{(n)} \sqrt{  \sum_{u_l \in \mathcal{U}_l } \left( S_{u_l, l}^{(n)}  \right)^2}&~{\leq}~ I_{th, 2}^{(n)}, ~~\forall n. \hspace*{2em} \label{eq:con-intf-2-relx_uncrt-CS}
 \end{eqnarray}
\setlength{\arraycolsep}{5pt}
\hrulefill
\vspace*{4pt}
\end{figure*}

Note that, as presented in Section \ref{subsec:uncert_set}, to tackle the uncertainty in channel gains, we have considered the worst-case approach, e.g., the estimation error is assumed to be bounded by a closed set (uncertainty set). Hence, from (\ref{eq:worst_bound_rate})-(\ref{eq:worst_bound_intf2}), under the worst-case channel uncertainties, the optimization problem $\mathbf{P3}$ can be rewritten as $\mathbf{P4}$, where $R_{u_l}$ is given by (\ref{eq:rate_uncet_final}). The transformed problem is a second-order cone program (SOCP) \cite[Chapter 4]{book-boyd} and the convexity of $\mathbf{P4}$ is conserved as shown in the following proposition.

\begin{figure*}[!t]
\normalsize
\begin{align} \label{eq:rate_uncet_final}
R_{u_l} &= \sum_{n =1}^N  \frac{1}{2} x_{u_l}^{(n)} B_{RB} 
\log_2 \Vast(  1 + \frac{\frac{S_{u_l, l}^{(n)}}{x_{u_l}^{(n)} }}{\sum\limits_{\substack{\forall u_j \in \mathcal{U}_j, \\ j \neq l, j \in \mathcal{L} }} \hspace{-0.4em} \bar{F}_{u_l, u_j, l}^{(n)} S_{u_j, j}^{(n)} + \xi_{1_{u_l}}^{(n)} \sqrt{\sum\limits_{\substack{\forall u_j \in \mathcal{U}_j, \\ j \neq l, j \in \mathcal{L} }} \hspace{-0.4em}  \big( S_{u_j, j}^{(n)} \big)^2 } + \tilde{\sigma}_{u_l}^{(n)}} \Vast)  \nonumber \\
& = \sum_{n =1}^N  \frac{1}{2} x_{u_l}^{(n)} B_{RB} 
\log_2 \Vast(  1 + \frac{\frac{S_{u_l, l}^{(n)}}{x_{u_l}^{(n)} } \bar{h}_{u_l, l, 1}^{(n)} }{\sum\limits_{\substack{\forall u_j \in \mathcal{U}_j, \\ j \neq l, j \in \mathcal{L} }} \hspace{-0.4em} \bar{g}_{u_j, l, 1}^{(n)} S_{u_j, j}^{(n)} +  \bar{h}_{u_l, l, 1}^{(n)} \xi_{1_{u_l}}^{(n)} \sqrt{\sum\limits_{\substack{\forall u_j \in \mathcal{U}_j, \\ j \neq l, j \in \mathcal{L} }} \hspace{-0.4em}  \big( S_{u_j, j}^{(n)} \big)^2 } + \sigma^{2}} \Vast)
\end{align}
\hrulefill
\vspace*{4pt}
\end{figure*}

\begin{proposition}
\label{prop:convex_uncert}
$\mathbf{P4}$ is a convex optimization problem.
\end{proposition} 
\begin{IEEEproof}
Using an argument similar to that in footnote \ref{ftnt:X2_constant}, the objective function of $\mathbf{P4}$ in (\ref{eq:relx_objfunc_uncrt_CS})  is concave. The constraints in (\ref{eq:con-bin}), (\ref{eq:con-pow-ue-relx}), (\ref{eq:x-s-relx}) are affine and the constraint in (\ref{eq:con-QoS-cue}) is convex. In addition, the additional square root term in the left hand side of the constraints  in (\ref{eq:con-pow-rel-relx_uncrt-CS}), (\ref{eq:con-intf-1-relx_uncrt-CS}), and (\ref{eq:con-intf-2-relx_uncrt-CS}) is the linear norm of
the vector of power variables $S_{u_l, l}^{(n)}$ with order $2$, which is convex \cite[Section 3.2.4]{book-boyd}. Therefore, the optimization problem $\mathbf{P4}$ is convex.
\end{IEEEproof}

$\mathbf{P4}$ is solvable 
using standard centralized algorithms such as interior point method. The joint RB and power allocation can be performed similar to \textbf{Algorithm \ref{alg:cent_soln}} and an upper bound for the solution can be obtained under channel uncertainty. It is worth noting that solving the above SOCP using interior point method incurs a complexity of $\mathcal{O} \left(\left( \overline{\overline{\mathbf{x}_{\boldsymbol l} }} + \overline{\overline{ \mathbf{P}_{\boldsymbol l} }} \right)^3 \right)$  at each relay node where $\overline{\overline{\mathbf{y}}}$ denotes the length of vector $\mathbf{y}$. Besides, the size of the optimization problem increases with the number of network nodes. Despite the fact that the solution from \textbf{Algorithm \ref{alg:cent_soln}} outputs the optimal data rate, considering very short scheduling period (e.g., $1$ millisecond in LTE-A network), it may not be feasible to solve the resource allocation problem centrally in practical networks. Therefore, in the following, we provide a low-complexity distributed solution based on matching theory. That is, without solving the resource allocation problem in a centralized manner using any relaxation technique (e.g., time-sharing strategy as described in the preceding section), we apply the method of  two-sided stable many-to-one matching \cite{matching_org_paper}. 

\section{Distributed Solution Approach Under Channel Uncertainty}
\label{sec:RA_SM}

The resource allocation approach using stable matching involves multiple decision-making agents, i.e., the available RBs and the UEs; and the solutions (i.e., matching between UE and RB) are produced by individual actions of the agents. The actions, i.e., matching requests and confirmations or rejections are determined by the given \textit{preference profiles}. That is, for both the RBs and the UEs, the lists of preferred matches over the opposite set are maintained. For each RB, the relay holds its preference list for the UEs. The matching outcome yields mutually beneficial assignments between RBs and UEs. \textit{Stability} in matching implies that, with regard to their initial preferences, neither RBs nor UEs have an incentive to alter the allocation.

\subsection{Concept of Matching}

A \textit{matching} (i.e., allocation) is given as an assignment of RBs to UEs forming the set of pairs $(u_l, n) \in \mathcal{U}_l \times \mathcal{N}$. Note that a UE can be allocated more than one RB to satisfy its data rate requirement; however, according to the constraint in (\ref{eq:con-bin}), one RB can be assigned to only one UE. This scheme corresponds to a \textit{many-to-one} matching in the theory of stable matching. More formally, we define the matching as follows \cite{matching_ubc}.

\begin{definition}
A matching $\mu_l$ for $\forall l \in \mathcal{L}$ is defined as a function, i.e.,  $\mu_l :  \mathcal{U}_l \cup \mathcal{N} \rightarrow \mathcal{U}_l \cup \mathcal{N} $ such that
\begin{enumerate} [i)]
\item $\mu_l(n) \in \mathcal{U}_l \cup \lbrace \varnothing \rbrace$ and $\overline{\overline{\mu_l(n)}} \in \lbrace 0,1 \rbrace$
\item $\mu_l(u_l) \in \mathcal{N}$ and $\overline{\overline{\mu_l(u_l)}} \in \lbrace 1, 2, \ldots, \kappa_{u_l} \rbrace$
\end{enumerate}
where the integer $\kappa_{u_l} \leq N$, $\mu_l(u_l) = n \Leftrightarrow \mu(n) = u_l$ for $ \forall n \in \mathcal{N}, \forall  u_l \in \mathcal{U}_l$ and $\overline{\overline{\mu_j(\cdot)}}$ denotes the cardinality of matching outcome $\mu_j(\cdot)$.
\end{definition}

The above definition implies that  $\mu_l$ is a one-to-one matching if the input to the function is an RB. On the other hand, $\mu_l$ is a one-to-many function, i.e., $\mu_l(u_l)$ is not unique if the input to the function is a UE. In order to satisfy the data rate requirement for each UE, we introduce the parameter $\kappa_{u_l}$ denoting the number RB(s) which are sufficient to satisfy the minimum rate requirement $Q_{u_l}$. Consequently, the constraint in (\ref{eq:con-QoS-cue}) is rewritten as $\displaystyle \sum_{n =1}^N    x_{u_l}^{(n)}    =  \kappa_{u_l}, ~ \forall  u_l$. Generally this parameter is referred to as \textit{quota} in the theory of matching~\cite{sm_def}. Each user $u_l$ will be subject to an acceptance quota $\kappa_{u_l}$ over RB(s) within the range $1\leq \kappa_{u_l}\leq N$ and allowed for matching to at most $\kappa_{u_l}$ RB(s). The outcome of the matching determines the RB allocation vector at each relay $l$, e.g., $\mu_l \equiv \mathbf{x}_{\boldsymbol l}$.

\subsection{Utility Matrix and Preference Profile}

Let us consider the utility matrix $\mathfrak{U}_l$ under the worst-case uncertainty, which denotes the achievable data rate for the UEs in different RBs, defined as follows:
\begin{equation}
\mathfrak{U}_l = \left[ 
\begin{smallmatrix} R_{1}^{(1)} & \cdots & R_{1}^{(N)}  \\ 
\vdots & \ddots & \vdots \\
R_{U_l}^{(1)} & \cdots & R_{U_l}^{(N)} 
\end{smallmatrix} 
\right]
\end{equation}
where $\mathfrak{U}_l[i,j]$ denotes the entry of $i$-th row  and $j$-th column in $\mathfrak{U}_l$, and $R_{u_l}^{(n)}$ is given by (\ref{eq:rate_sm_uncrt}).
\begin{figure*}[!t]
\normalsize
\begin{equation} \label{eq:rate_sm_uncrt}
R_{u_l}^{(n)} = \frac{1}{2} B_{RB} 
\log_2 \Vast(  1 + \frac{P_{u_l, l}^{(n)} \bar{h}_{u_l, l, 1}^{(n)} }{\sum\limits_{\substack{\forall u_j \in \mathcal{U}_j, \\ j \neq l, j \in \mathcal{L} }} \hspace{-0.4em} x_{u_j}^{(n)} \bar{g}_{u_j, l, 1}^{(n)} P_{u_j, j}^{(n)} +  \bar{h}_{u_l, l, 1}^{(n)} \xi_{1_{u_l}}^{(n)} \sqrt{\sum\limits_{\substack{\forall u_j \in \mathcal{U}_j, \\ j \neq l, j \in \mathcal{L} }} \hspace{-0.4em}  \big( x_{u_j}^{(n)} P_{u_j, j}^{(n)} \big)^2 } + \sigma^{2}} \Vast)
\end{equation}
\hrulefill
\vspace*{4pt}
\end{figure*}
Each  of the UEs and RBs holds a list of preferred matches where a preference relation can be defined as follows \cite[Chapter 2]{pref_profile_book}.

\begin{definition}
Let $\succeq$ be a binary relation on any arbitrary set $\Xi$. The binary relation $\succeq$ is \textbf{complete} if for $\forall i, j \in \Xi$, either $i \succeq j$ or $j \succeq i$ or both. A binary relation is \textbf{transitive} if $i \succeq j$ and $j \succeq k$ implies that $i \succeq k$ for $ \forall k \in \Xi$. The binary relation $\succeq$ is a (weak) \textbf{preference} relation if it is complete and transitive.
\end{definition}

The preference profile of a UE $u_l \in \mathcal{U}_l$ over the set of available RBs $\mathcal{N}$ is defined as a vector of linear order $\boldsymbol{\mathscr{P}}_{u_l}(\mathcal{N}) = \mathfrak{U}_l[u_l, i]_{i\in \mathcal{N}}$. The UE $u_l$ prefers RB $n_1$ to $n_2$ if $n_1 \succeq n_2$, and consequently, $\mathfrak{U}_l[u_l, n_1] > \mathfrak{U}_l[u_l, n_2]$. Likewise, the preference profile of an RB $n \in \mathcal{N}$ is given by $\boldsymbol{\mathscr{P}}_{n}(\mathcal{U}_l) = \mathfrak{U}_l[j, n]_{j\in \mathcal{U}_l}$.

\subsection{Algorithm for Resource Allocation}

Based on the discussions in the previous section, we utilize an improved version of matching algorithm (adapted from \cite[Chapter 1.2]{matching_thesis}) to allocate the RBs. The allocation subroutine, as illustrated in \textbf{Algorithm \ref{alg:rb_alloc}}, executes as follows. While an RB $n$ is unmatched (i.e., unallocated) and has a non-empty preference list, the RB is temporarily assigned to its first preference over UEs, i.e., $u_l$. If the allocation does not exceed $\kappa_{u_l}$, the allocation will persist. Otherwise, the worst preferred RB from  $u_l$'s matching will be removed even though it was previously allocated. The iterations are repeated until there are unallocated pairs of RB and UE. The iterative process dynamically updates the preference lists and hence leads to a stable matching.  

\begin{algorithm} [!t]
\AtBeginEnvironment{algorithmic}{\small} 
\caption{RB allocation using stable matching}
\label{alg:rb_alloc}
\begin{algorithmic}[1]   
\renewcommand{\algorithmicrequire}{\textbf{Input:}}
\renewcommand{\algorithmicensure}{\textbf{Output:}}
\renewcommand{\algorithmicforall}{\textbf{for each}}
\renewcommand{\algorithmiccomment}[1]{\textit{/* #1 */}}

\REQUIRE The preference profiles $\boldsymbol{\mathscr{P}}_{u_l}(\mathcal{N})$, $\boldsymbol{\mathscr{P}}_{n}(\mathcal{U}_l)$; $\forall u_l \in \mathcal{U}_l, n \in \mathcal{N}$.

\ENSURE The RB allocation vector $\mathbf{x}_{\boldsymbol l}$.

\STATE Initialize $\mathbf{x}_{\boldsymbol l} := \mathbf{0}$.

\WHILE{ ($\exists u_l$ with  $\sum\limits_{n=1}^{N}x_{u_l}^{(n)} < \kappa_{u_l}$ ) \OR ($\exists n$ with $x_{u_l}^{(n)} = 0, \forall u_l \in \mathcal{U}_l$ \AND $\boldsymbol{\mathscr{P}}_{n}(\mathcal{U}_l) \neq \varnothing$) }
\STATE $u_{mp} :=$ most preferred UE from the profile $\boldsymbol{\mathscr{P}}_{n}(\mathcal{U}_l)$. 
\STATE Set $x_{u_{mp}}^{(n)} := 1$.  ~\COMMENT{\footnotesize Temporarily allocate the RB}
\IF{$\displaystyle \sum_{j = 1}^{N} x_{u_{mp}}^{(j)} > \kappa_{u_{mp}}$}
\STATE $n_{lp} :=$ least preferred resource allocated to $u_{mp}$.
\STATE Set $x_{u_{mp}}^{(n_{lp})} := 0$. ~\COMMENT{\footnotesize Revoke allocation due to quota violation}
\ENDIF 

\IF{$\displaystyle \sum_{j = 1}^{N} x_{u_{mp}}^{(j)} = \kappa_{u_{mp}}$}
\STATE $n_{lp} :=$ least preferred resource allocated to $u_{mp}$.
\STATE { \footnotesize \textit{/* Update preference profiles */} }
\FORALL {successor $\hat{n}_{lp}$ of $n_{lp}$ on profile $\boldsymbol{\mathscr{P}}_{u_{mp}}(\mathcal{N})$} 

\STATE remove $\hat{n}_{lp}$ from $\boldsymbol{\mathscr{P}}_{u_{mp}}(\mathcal{N})$.
\STATE remove $u_{mp}$ from $\boldsymbol{\mathscr{P}}_{\hat{n}_{lp}}(\mathcal{U}_l)$. 
\ENDFOR

\ENDIF 

\ENDWHILE

\end{algorithmic}
\end{algorithm}

Once the optimal RB allocation is obtained, the transmit power of the UEs on assigned RB(s) is obtained as follows. We couple the classical generalized distributed constrained power control scheme (GDCPC) \cite{power_gdpc} with an autonomous power control method \cite{auto_pow_cr} which considers the data rate requirements of UEs while protecting other receiving nodes from interference. More specifically, at each iteration $t$, the transmission power for each allocated RB is updated as follows:
\begin{align} \label{eq:pow_alloc}
P_{u_l,l}^{(n)}(t) = \begin{cases}
 \Lambda(t-1), \quad  \text{if $\Lambda(t-1) \leq \hat{P}_{u_l}^{{(n)}^{max}}$} \vspace{0.5em}\\ 
 \hat{P}_{u_l,l}^{(n)}, \quad \quad ~\text{otherwise} 
\end{cases}
\end{align}
where 
\vspace{-1em}
\begin{align}
\Lambda(t-1) =  \tfrac{2^{Q_{u_l}  } -1 }{2^{R_{u_l}(t-1) } - 1} P_{u_l,l}^{(n)}(t-1) \hspace{3.0em} \\
\hat{P}_{u_l}^{{(n)}^{max}} = \min \left( \tfrac{ P_{u_l}^{max} } { \sum\limits_{n=1}^{N} x_{u_l}^{(n)} }, \tfrac{P_{l}^{max}}{ \left( \bar{H}_{u_l, l}^{(n)} + {\xi_2}_{u_l} \right) \sum\limits_{u_l \in \mathcal{U}_l} \sum\limits_{n=1}^{N}  x_{u_l}^{(n)} }   \right)
\end{align}
and $ \hat{P}_{u_l,l}^{(n)}$ is obtained as 
\begin{align} \label{eq:power_p_tilde}
 \hat{P}_{u_l,l}^{(n)} = \min \left( \tilde{P}_{u_l,l}^{(n)}, ~ \min \left( \hat{P}_{u_l}^{{(n)}^{max}}, \varpi_{u_l,l}^{(n)} \right) \right).
\end{align}
In (\ref{eq:power_p_tilde}), the parameter $\tilde{P}_{u_l,l}^{(n)}$ is chosen arbitrarily within the range of $0 \leq \tilde{P}_{u_l,l}^{(n)} \leq \hat{P}_{u_l}^{{(n)}^{max}}$ and $\varpi_{u_l,l}^{(n)} $ is given by
\begin{equation}
\varpi_{u_l,l}^{(n)} =  \min \left( \tfrac{  I_{th, 1}^{(n)} } { \bar{g}_{{u_l^*}, l, 1}^{(n)} + \xi_{3_{u_l}}^{(n)}  },  \tfrac{ I_{th, 2}^{(n)} }{  \bar{H}_{u_l, l}^{(n)} \bar{g}_{l, {u_l^*}, 2}^{(n)} + \xi_{4_{u_l}}^{(n)}  }    \right).
\end{equation}

Based on the RB allocation, the relay informs the parameter $\hat{P}_{u_l}^{{(n)}^{max}}$ and each UE updates its transmit power in a distributed manner using (\ref{eq:pow_alloc}).  Each relay independently performs resource allocation and allocates resources to corresponding associated UEs. The joint RB and power allocation algorithm is given in \textbf{Algorithm \ref{alg:joint_alloc}}. 

\begin{algorithm} [!t]
\AtBeginEnvironment{algorithmic}{\small} 
\caption{Joint RB and power allocation algorithm}
\label{alg:joint_alloc}
\begin{algorithmic}[1]   
\renewcommand{\algorithmicrequire}{\textbf{Input:}}
\renewcommand{\algorithmicensure}{\textbf{Output:}}
\renewcommand{\algorithmicforall}{\textbf{for each}}
\renewcommand{\algorithmiccomment}[1]{\textit{/* #1 */}}

\renewcommand{\algorithmicensure}{\textbf{Phase I: Initialization}}
\ENSURE

\STATE  Each relay  $l \in \mathcal{L}$ estimates the nominal CQI values from previous time slot and determines reference gains $\bar{g}_{u_l^*, l, 1}^{(n)}$ and  $\bar{g}_{l, u_l^*, 2}^{(n)}, \forall u_l, n$.

\STATE Initialize $t:=0$, $P_{u_l,l}^{(n)} := \frac{P_{u_l}^{max}}{N}$ $\forall u_l, n$ and $\mathfrak{U}_l$ based on CQI estimates.

\renewcommand{\algorithmicensure}{\textbf{Phase II: Update}}
\ENSURE

\FORALL {relay $l \in \mathcal{L}$ }

\REPEAT 

\STATE Update $t:= t + 1$.

\STATE Build the preference profile $\boldsymbol{\mathscr{P}}_{n}(\mathcal{U}_l)$ for each RB $n \in \mathcal{N}$ based on utility matrix and inform corresponding entries of $\mathfrak{U}_l$ to UEs. 

\STATE Each UE $u_l \in \mathcal{U}_l$ builds the preference profile $\boldsymbol{\mathscr{P}}_{u_l}(\mathcal{N})$.

\STATE Obtain RB allocation vector using \textbf{Algorithm \ref{alg:rb_alloc}}.

\STATE Update the transmission power using (\ref{eq:pow_alloc}) for $\forall u_l, n$ and update the utility matrix $\mathfrak{U}_l$. 

\STATE Inform the allocation variables $\mathbf{x}_{\boldsymbol l}$,  $\mathbf{P}_{\boldsymbol l}$ to each relay $j \neq l, j \in \mathcal{L}$ and calculate the achievable data rate based on current allocation as  $\displaystyle R_l(t) := \sum_{u_l \in \mathcal{U}_l} R_{u_l}(t)$.

\UNTIL data rate not maximized \AND $t < T_{max}$.

\ENDFOR

\renewcommand{\algorithmicensure}{\textbf{Phase III: Allocation}}
\ENSURE

\STATE  For each relay, allocate resources (i.e., RB and transmit power) to the associated UEs.

\end{algorithmic}
\end{algorithm}

\section{Analysis of the Proposed Solution} \label{sec:analysis_sm}

In the following, we analyze the performance of our proposed distributed resource allocation approach under  bounded channel uncertainty. More specifically, we analyze the stability, optimality, and uniqueness of the solution, and its computational complexity.

\subsection{Stability}

\begin{definition}  \label{def:rational}
\begin{enumerate}[(a)]
\item The pair of UE and RB $(u_l, n)$ in $\mathcal{U}_l \times \mathcal{N}$ is \textbf{acceptable} if $u_l$ and $n$ prefer each other (to be matched) to being remain unmatched.
\item A matching $\mu_l$ is called \textbf{individually rational} if no agent  (i.e., UE or RB) $\tilde{\jmath}$ prefers to remain unmatched to $\mu(\tilde{\jmath})$.
\end{enumerate}

\end{definition}

\begin{definition} \label{def:blocking}
A matching $\mu_l$ is \textbf{blocked} by a pair of agents $(i,j)$ if they each prefer each other to the matching they obtain by $\mu_l$, i.e.,  $i \succeq \mu_l(j)$ and $j \succeq \mu_l(i)$.
\end{definition}

From \textbf{Definition \ref{def:rational}}, \textbf{\ref{def:blocking}}, the matching $\mu_l$ is blocked by RB $n$ and UE $u_l$ if $n$ prefers $u_l$ to $\mu_l(n)$ and either  \begin{inparaenum}[\itshape i\upshape)]
\item $u_l$ prefers $n$ to some $\hat{n} \in \mu_l(u_l)$, or
\item $\overline{\overline{\mu_l(u_l)}} < \kappa_{u_l}$ and $n$ is acceptable to $u_l$. 
\end{inparaenum}
Using the above definitions, the stability of matching can be defined as follows \cite[Chapter 5]{matchingbook_two}.

\begin{definition}  \label{def:blocking2}
A matching $\mu_l$ is \textbf{stable} if it is individually rational and there is no pair $(u_l, n)$ in the set of acceptable pairs such that $u_l$ prefers $n$ to  $\mu_l(u_l)$ and $n$ prefers $u_l$ to  $\mu_l(n)$, i.e., not blocked by any pair of agents. 

\end{definition}

\begin{proposition}
\label{prop:stability}
The assignment performed in {\normalfont \textbf{Algorithm \ref{alg:rb_alloc}}} abides by the preferences of the UEs and RBs and it leads to a stable allocation.
\end{proposition} 
\begin{IEEEproof}
See \textbf{Appendix \ref{appsec:stability}}.
\end{IEEEproof}

\vspace{0.3cm}
Note that the allocation of RBs is stable at each iteration of \textbf{Algorithm \ref{alg:joint_alloc}}. Since after  evaluation of the utility, the preference profile of UEs and RBs are updated and the routine for  RB allocation  is repeated, a stable allocation is obtained.

\subsection{Uniqueness}

\begin{proposition}
\label{prop:uniq}
If there are sufficient number of RBs (i.e., $N \geq U_l$), and the preference lists of all UEs and RBs are determined by the $U_l \times N$ utility matrix $\mathfrak{U}_l$ whose entries are all different and obtained from given uncertainty bound, then there is a unique stable matching.
\end{proposition} 
\begin{IEEEproof}
See \textbf{Appendix \ref{appsec:uniq}}.
\end{IEEEproof}

\subsection{Optimality and Performance Bound}

\begin{definition} \label{def:pareto}
A matching $\mu_l$ is weak Pareto optimal if there is no other matching $ \widehat{\mu}_l$ that can achieve a better sum-rate, i.e., $\widehat{\mu}_l(\cdot) \geq \mu_l(\cdot) $, where the inequality is component-wise and strict for one user. 
\end{definition}

\begin{proposition}
\label{prop:optimality}
The proposed resource allocation algorithm is weak Pareto optimal under bounded channel uncertainty.
\end{proposition} 
\begin{IEEEproof}
See \textbf{Appendix \ref{appsec:optimality}}.
\end{IEEEproof}

\begin{corollary}
\label{cor:algo_bound}
Since ${\mathbf{x}_{\boldsymbol l}}^*$ satisfies the binary constraint in (\ref{eq:bin_con}), and the optimal allocation  $\left({\mathbf{x}_{\boldsymbol l}}^*, {\mathbf{P}_{\boldsymbol l}}^*\right)$ satisfies all the constraints in the optimization problem $\mathbf{P4}$, for a sufficient number of available RBs, the data rate obtained by {\normalfont \textbf{Algorithm \ref{alg:joint_alloc}}} gives a lower bound of the solution under channel uncertainty.
\end{corollary}

\subsection{Convergence and Computational Complexity}

\begin{proposition}
\label{prop:time}
The subroutine for RB allocation  terminates after some finite number of steps $T^\prime$.
\end{proposition} 
\begin{IEEEproof}
Let the finite set $\tilde{\mathcal{X}}$ represent all possible combinations of UE-RB matching where each element  $\tilde{x}_{i}^{(j)} \in \tilde{\mathcal{X}}$ denotes that RB $j$ is allocated to UE $i$. Since no UE rejects the same RB more than once (see line $7$ in \textbf{Algorithm \ref{alg:rb_alloc}}), the finiteness of the set $\tilde{\mathcal{X}}$ ensures the termination of RB allocation subroutine in finite number of steps.
\end{IEEEproof}

Note that the distributed approach replaces the optimization routine in the centralized approach with the matching algorithm (\textbf{Algorithm 2}). Since by \textbf{Proposition \ref{prop:time}} we show matching algorithm terminates after finite number of iterations, ultimately \textbf{Algorithm 3} will end up with a (local) Pareto optimal solution after some finite number of iterations.

\mathchardef\mhyphen="2D
In line $6\mhyphen7$ of \textbf{Algorithm \ref{alg:joint_alloc}}, the complexity to output the ordered set of preference profiles for the RBs using any standard sorting algorithm is  $\mathcal{O}\left( N U_l \log U_l \right)$ and for each UE, the complexity to build the preference profile is $\mathcal{O}\left( N \log N \right)$. Let $\beta = \displaystyle \sum_{u_l = 1 }^{U_l} \overline{\overline{\boldsymbol{\mathscr{P}}_{u_l}(\mathcal{N})}} + \sum_{n = 1}^{N} \overline{\overline{\boldsymbol{\mathscr{P}}_{n}(\mathcal{U}_l)}}  = 2N U_l$ be the total length of input preferences in \textbf{Algorithm \ref{alg:rb_alloc}}, where $\overline{\overline{\boldsymbol{\mathscr{P}}_j(\cdot)}}$ denotes the length of the profile vector $\boldsymbol{\mathscr{P}}_j(\cdot)$. From \textbf{Proposition \ref{prop:time}} and \cite[Chapter 1]{matching_thesis} it can be shown that, if implemented with suitable data structures, the time complexity of RB allocation subroutine is linear in the size of input preference profiles, i.e., $\mathcal{O}(\beta) \approx \mathcal{O}\left( NU_l \right)$. Since \textbf{Phase II} of \textbf{Algorithm \ref{alg:joint_alloc}} runs at most fixed $T_{max}$ iterations, at each relay node $l$, the complexity of the proposed solution is linear in $\overline{\overline{\mathcal{N}}}$ and $\overline{\overline{\mathcal{U}_l}}$.

\subsection{Signalling Over Control Channels}

Assuming that the relays obtain the CQI  prior to resource allocation, 
the centralized approach  does not require any exchange of information between  a relay node and the associated UEs to perform resource allocation. However,
in the distributed approach, the relay node and the UEs need to exchange information to update the preference profiles and transmit power. In both the approaches, the relay nodes need to exchange the allocation variables among themselves (e.g., over X2 interface) in order to calculate the interference levels at the receiving nodes.

In the distributed approach, the exchange of information between a UE and the relay node during execution of the resource allocation algorithm can be mapped onto the standard LTE-A scheduling control messages. For scheduling in LTE-A networks, the exchanges of messages over control channels  are as follows \cite{signalling_3gpp}. The UEs will periodically sense the physical uplink control channel (PUCCH)  by transmitting known sequences as sounding reference signals (SRS). 
When data is available for uplink transmission, the UE sends the scheduling request (SR) over PUCCH. The relay, in turn, uses the scheduling grant (SG) over physical downlink control channel (PDCCH) to allocate the appropriate RB(s) to the UE. Once the allocation of RB(s) is received, the UE regularly sends buffer status report (BSR) using PUCCH in order to update the resource requirement, and in response, the relay sends the acknowledgment (ACK) over the physical hybrid-ARQ indicator channel (PHICH). Given the above scenario, the UEs may provide the preference profile $\boldsymbol{\mathscr{P}}_{u_l}(\mathcal{N})$ with the SR and BSR messages. The relays may provide the corresponding values in the utility matrix, e.g., $\boldsymbol{\hat{\mathfrak{u}}_{u_l, l}} = \mathfrak{U}_l[u_l, j]_{j = 1, \cdots, N}$ and inform the parameter $\hat{P}_{u_l}^{{(n)}^{max}}$ using SG and ACK messages. Once the RB and power allocation is performed, the relays multicast the allocation information over X2 interface. 

In what follows, we analyze signalling overhead for our proposed solution. For a sufficient number of available RBs, we consider two cases: \begin{inparaenum}[\itshape a\upshape)]
\item number of available RBs at each relay $l$ is equal to the number of UEs (e.g., $N = U_l$ and $\kappa_{u_l} = 1,~ \forall u_l$); and
\item the number of RBs is greater than the number of UEs (e.g., $N > U_l$).
\end{inparaenum} 
For the first case, once \textbf{Algorithm 2} terminates, all RBs are allocated to the UEs. This is because, by the definition of individual rationality (see \textbf{Definition 3}), none of the agents (i.e., UE or RB) wants to remain unallocated. Hence, at the end of any iteration $\hat{t}$ of \textbf{Algorithm 2}, there are $N - \hat{t}$ unallocated RBs at each relay $l$. Therefore, the maximum number of iterations, say $\widehat{T}_{max}$ are required when all the RBs are allocated, e.g.,   $N - \widehat{T}_{max} = 0$, and therefore, $\widehat{T}_{max} = N$. Since at each iteration $\hat{t}$, $N - \hat{t} + 1$ messages are exchanged, the total number of messages exchanged in \textbf{Algorithm 2} can be quantified as
\begin{equation}
\Omega_l = \sum_{i=1}^{\widehat{T}_{max}} \left( N - i + 1 \right) = \frac{N(N+1)}{2}.
\end{equation} 
In \textbf{Algorithm 3}, each relay $l$ exchanges the allocation parameters (e.g., $\mathbf{x}_{\boldsymbol l}$,  $\mathbf{P}_{\boldsymbol l}$) over X2 interface. If \textbf{Algorithm 3} executes $T < T_{max}$ iterations, the overall signalling overhead (e.g., number of messages exchanged)  is given by
\begin{equation} \label{eq:sig-1}
\Omega_{l}^{max} = T (\Omega_l + 1) = T \times \frac{N^2+N+2}{2}.
\end{equation} 
Likewise, for the second case (e.g.,  when $N > U_l$), \textbf{Algorithm 2}  terminates when there are no unallocated UEs with less than their quota requirement (e.g., $\sum\limits_{n=1}^{N}x_{u_l}^{(n)} < \kappa_{u_l}$). Hence the maximum number of iterations $\widehat{T}_{max} = U_l$, and the numnber of messages  exchanged in \textbf{Algorithm 2} for the second case is given by
\begin{align}
\Omega_l &= \sum_{i=1}^{\widehat{T}_{max}} \left( N - i + 1 \right) = \sum_{i=1}^{U_l} N - i + 1 \nonumber \\
&= (N+1)U_l - \frac{U_l(U_l+1)}{2}.
\end{align}
Therefore, the overall signalling overhead for the second case can be expressed as follows:
\begin{equation} \label{eq:sig-2}
\Omega_{l}^{max} = T \times \frac{2(N+1)U_l - U_l(U_l+1) + 2}{2}.
\end{equation}
As can be seen from (\ref{eq:sig-1}) and (\ref{eq:sig-2}), the overhead of signalling increases with the number of UEs and the available RBs. However, the proposed distributed approach has significantly lower computational complexity than the centralized approach (linear compared to cubic) and it offers performance improvement over the existing solutions (see Section \ref{sec:res}).

\section{Performance Evaluation} \label{sec:numerical_results}

\subsection{Simulation Setup}

We develop a discrete-time simulator in MATLAB and evaluate the performance of our proposed solution. We simulate a single three-sectored cell in a rectangular area of $700~\text{m} \times 700~\text{m}$, where the eNB is located in the center of the cell and three relays are deployed, i.e., one relay in each sector. The CUEs are uniformly distributed within the relay cell. The D2D UEs are located according to the clustered distribution model \cite{d2d_cluster_dist_model}. In particular, the D2D transmitters are uniformly distributed over a radius $D_{r,d}$; and the D2D receivers are distributed uniformly in the perimeter of the circle with radius $D_{d,d}$ centered at the corresponding D2D transmitter (Fig. 2 in \cite{d2d_our_robust}).  Both $D_{r,d}$ and $D_{d,d}$ are varied as simulation parameters and
the values are specified in the corresponding figures. The simulation results are averaged over $200$ network realizations of user locations and channel gains. We consider a snapshot model and all the network parameters are assumed to remain unchanged during a simulation run. For propagation modeling, we consider distance-dependent path-loss, shadow fading, and multi-path Rayleigh fading (see Section VII-A in \cite{d2d_our_robust}). 

We measure the uncertainty in channel gains as percentages and assume similar uncertainty bounds in the CQI parameters for all the UEs. For example, uncertainty bound $\xi = \xi_{1_{u_l}}^{(n)} = \xi_{2_{l}} = \xi_{3_{u_l}}^{(n)} =  \xi_{4_{u_l}}^{(n)} = 0.25$ refers that uncertainty (e.g., estimation error) in the CQI parameters for $\forall u_l, n, l$ is not more than $25\%$ of their nominal values. The simulation parameters are similar to those  in Table~II in \cite{d2d_our_robust}. 

\subsection{Results} \label{sec:res}

A summary of the observations from the performance evaluation results is provided in Table \ref{tab:obsrv}.

\begin{table}[!t]
\caption{Summery of performance results}
\label{tab:obsrv}
\centering
\begin{tabular}{l p{21em}|l}
\hline
& \hspace*{7.0em} \bfseries  Observations & \bfseries Ref. figure(s)\\
\hline\hline
$\bullet$ \hspace{-1em} & The proposed distributed solution converges to a stable data rate within a few  iterations and performs close to the optimal data rate with significantly less computational complexity. & Figs. \ref{fig:convergence_sm}-\ref{fig:ub_ue} \\
$\bullet$ \hspace{-1em} & As the distance between D2D peers increases, the data rate for direct communication decreases. In such cases relaying of D2D traffic can improve the end-to-end data rate between D2D peers.  & Fig. \ref{fig:rate_vs_d2d_dist}  \\
$\bullet$ \hspace{-1em} &  After a distance threshold, relay-aided D2D communication provides considerable gain in terms of the achievable data rate for the D2D UEs. Even for a relatively large distance between the relay node and a D2D UE, relaying can provide a better data rate compared to direct communication for distant D2D peers. There is also a trade-off between achievable data rate and robustness against channel uncertainty. & Figs. \ref{fig:gain_vs_d2d_dist}-\ref{fig:rel_var_sm}  \\
\hline
\end{tabular}
\end{table}

\begin{figure}[!t]
\centering
\includegraphics[scale = 0.50]{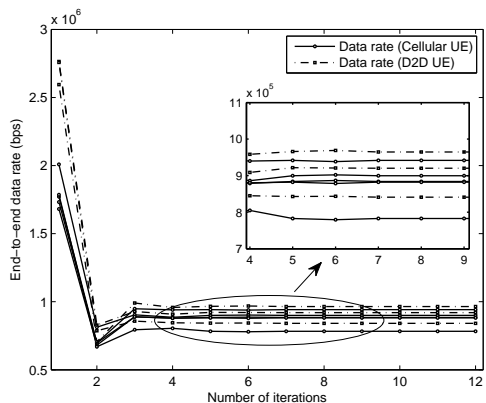}
\caption{Convergence of the proposed solution where the number of CUEs and D2D UEs served by each relay node is $5$ and $3$, receptively (e.g., $|\mathcal{U}_l| = 8$). $D_{r,d}$ and $D_{d,d}$ are set to $50~ {\rm m}$, and uncertainty in CQI parameters is assumed to be not more than $25\%$.}
\label{fig:convergence_sm}
\end{figure}

\subsubsection{Convergence and goodness of the solution}
In Fig. \ref{fig:convergence_sm}, we show the convergence behavior of our proposed distributed algorithm. In particular, we plot the average achievable data rate for the UEs in different network realizations versus the number of iterations. The algorithm starts with uniform power allocation over RBs, which provides a higher data rate at the first iteration; however, it  may cause severe interference to other receiving nodes. As the algorithm executes, the allocations of RB and power are updated  considering the interference  threshold and data rate constraints. From this figure it can be observed that the solution converges to a stable data rate very quickly (e.g., in less than $10$ iterations).

\begin{figure}[ht]
\centering
\subfigure[]{%
\includegraphics[scale = 0.50]{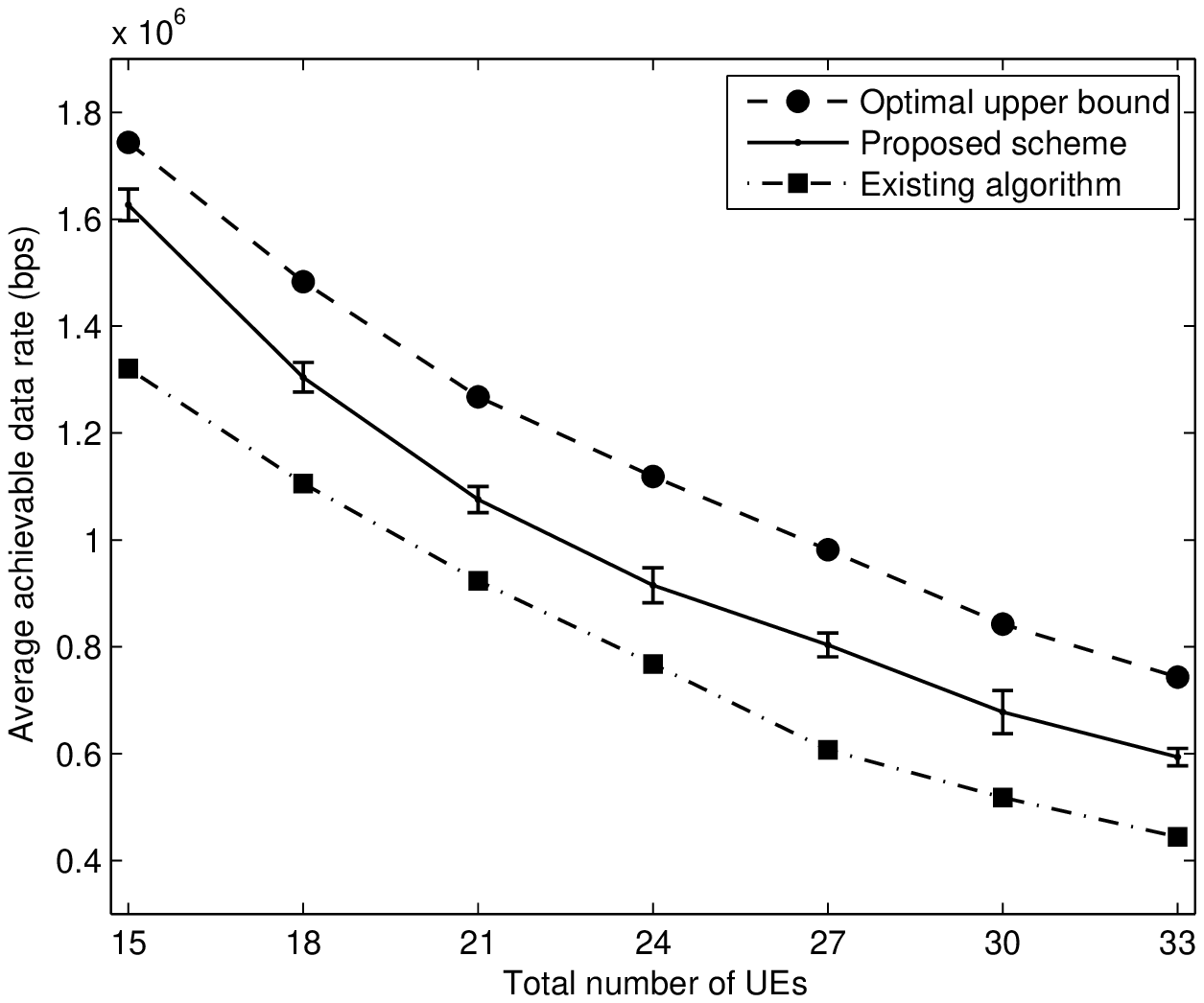}
\label{fig:compare_ub}}
\quad
\subfigure[]{%
\includegraphics[scale = 0.50]{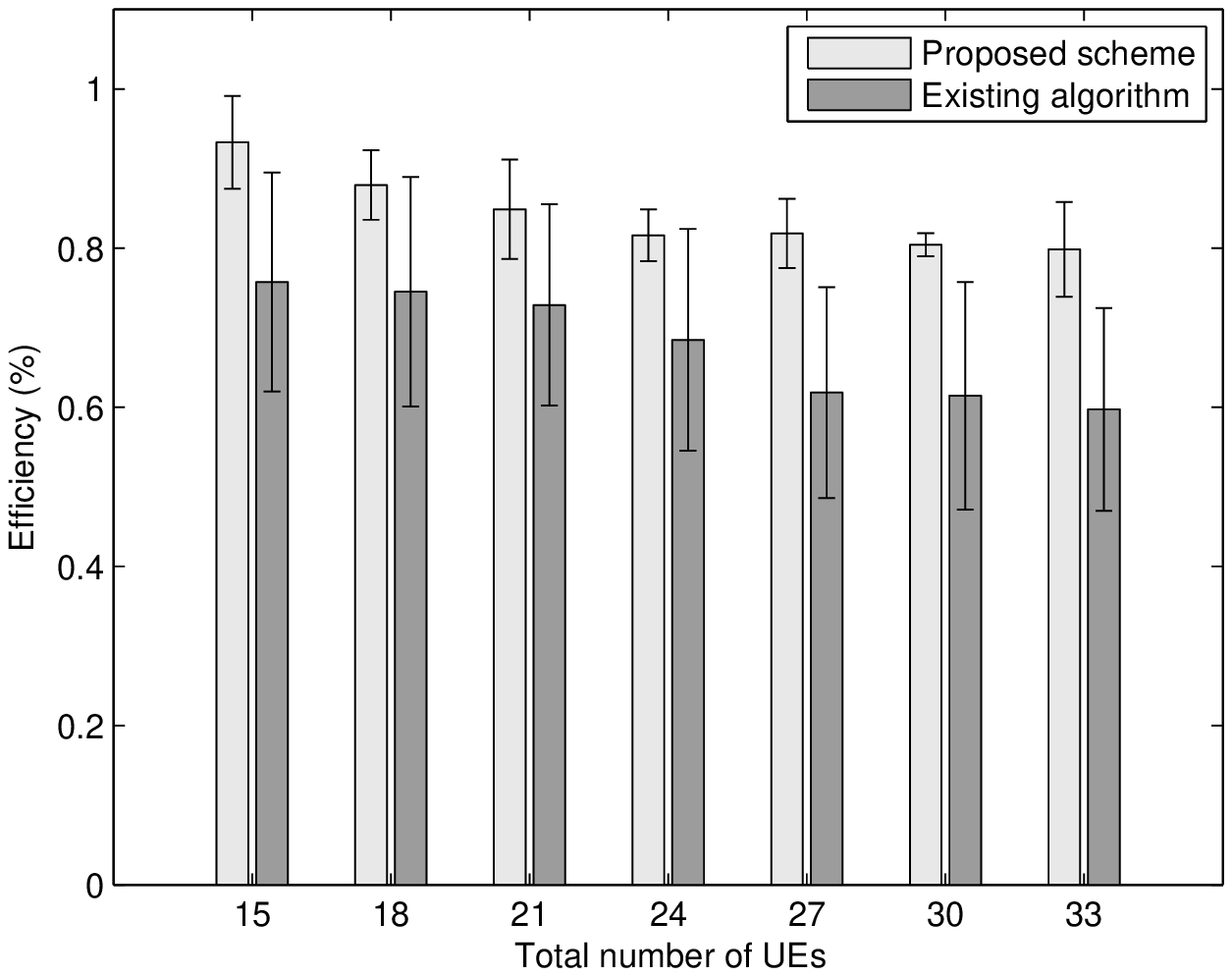}
\label{fig:efficiency}}
\caption{(a) Average achievable data rate for optimal upper bound, distributed stable matching and existing algorithm. (b) Efficiency of the proposed solution and the existing algorithm. Total number of UEs (i.e., $C + D $) are varied from $9+6 = 15$ to $15 + 18 = 33$. $D_{r,d}$ and $D_{d,d}$ are assumed to be $50~ {\rm m}$.}
\label{fig:ub_ue}
\end{figure}

We compare the performance of our proposed scheme with a dual-decomposition based suboptimal resource allocation scheme proposed in \cite{hetnet_dual_comp}. We refer to this scheme as \textit{existing algorithm}. In this scheme, the relay node allocates RBs considering the data rate requirement and the transmit power is updated in an iterative manner by updating the Lagrange dual variables. For details refer to \cite[Algorithm 2]{hetnet_dual_comp}. The complexity of this algorithm is of $\mathcal{O}\left( N U_l \log N +  N \log U_l + \Delta \right)$, where $\Delta$ denotes the number of iterations it takes for the power allocation vector to converge \cite{hetnet_dual_comp}.

In Fig. \ref{fig:compare_ub}, we show the performances of the proposed distributed scheme and the existing algorithm, and the upper bound of the optimal solution which can be obtained  in a centralized manner using \textbf{Algorithm \ref{alg:cent_soln}}. We use the MATLAB optimization toolbox to obtain this upper bound. We plot the average achievable data rate for the UEs versus the total number of UEs. The average data rate is given by $R_{\rm avg} =  \frac{ \displaystyle \sum_{u \in \lbrace\mathcal{C} \cup \mathcal{D} \rbrace} R_{u}^{\rm ach}}{C+D}$, where $R_{u}^{\rm ach}$ is the achievable data rate for UE $u$. Note that, for a given number of RBs, increasing the number of UEs decreases the data rate.  
Recall that, the complexity of both the proposed and reference schemes is linear with the number of RBs and UEs; and for the optimal solution, the complexity is cubic to the number of RBs and UEs. As can be seen from this figure, the proposed approach outperforms the existing algorithm and performs close to the optimal solution.

In order to obtain more insights into the performance, in Fig \ref{fig:efficiency}, we plot the efficiency of the proposed scheme and existing algorithm for different number of UEs. Similar to \cite[Chapter 3]{effi_zuhan_book}, we measure the efficiency as  $\eta_{(\cdot)} = \frac{R_{(\cdot)}}{R_{\rm optm}} $,  where $R_{\rm optm}$ is the network sum-rate for optimal solution. The parameters $R_{\rm prop}$ and $R_{\rm exst}$ denote the data rate for the proposed  and existing schemes, respectively, which are used to calculate the corresponding efficiency metric  $\eta_{\rm prop}$ and $\eta_{\rm exst}$. The closer the value of $\eta_{(\cdot)}$ to $1$, the nearer the solution is to the optimal solution. 
Clearly, the efficiency of the existing algorithm is lower compared to the proposed scheme. From the figure we observe that even in a dense network scenario (i.e., $C + D = 15 + 18 = 33$) the proposed scheme performs $80\%$ close to the optimal solution (compared to $60\%$ for the existing algorithm); however, with much less computational complexity.

\subsubsection{Impact of relaying}

We compare the performance of the proposed method for relay-aided D2D communication with a conventional underlay D2D communication scheme. In this scheme\cite{zul-d2d}, an RB allocated to CUE can be shared with at most one D2D link. The D2D UEs share the same RB(s) (allocated to a CUE using \textbf{Algorithm \ref{alg:joint_alloc}}) and communicate directly with their peers \textit{without} a relay if the data rate requirements for both the CUEs and D2D UEs are satisfied; otherwise, the D2D UEs refrain from transmitting. We refer to this underlay D2D communication scheme  \cite{zul-d2d} as the \textit{reference scheme}. 
Notice that this conventional (e.g., direct) D2D communication approach can save half of the RBs. Therefore, as mentioned in footnote \ref{ftn:direct_d2d_rate}, data rate of $u_l$ in the reference scheme is given by $R_{u_l}^{\rm ref} = \sum\limits_{n =1}^N  x_{u_l}^{(n)} B_{RB} \log_2 \left( 1 + P_{u_l}^{(n)} \tilde{\gamma}_{u_l}^{(n)} \right)$. On the contrary, data rate of $u_l$ in the proposed relay-aided approach is given by $R_{u_l}^{\rm prop} = \sum\limits_{n =1}^N   \tfrac{1}{2}  x_{u_l}^{(n)}   B_{RB} \log_2   \left(  1 +   P_{u_l, l}^{(n)} \gamma_{u_l, l, 1}^{(n)}   \right) $.

\begin{figure}[!t]
\centering
\includegraphics[scale = 0.50]{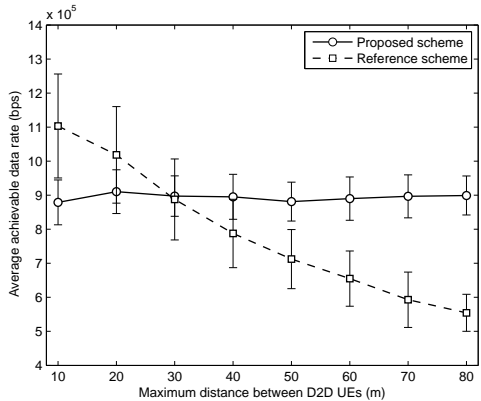}
\caption{Gain in average achievable data rate with varying distance between D2D peers using a setup similar to that for Fig. \ref{fig:convergence_sm}. The reference scheme is an underlay D2D communication approach proposed in \cite{zul-d2d}.}
\label{fig:rate_vs_d2d_dist}
\end{figure}
\begin{figure}[!t]
\centering
\includegraphics[scale = 0.50]{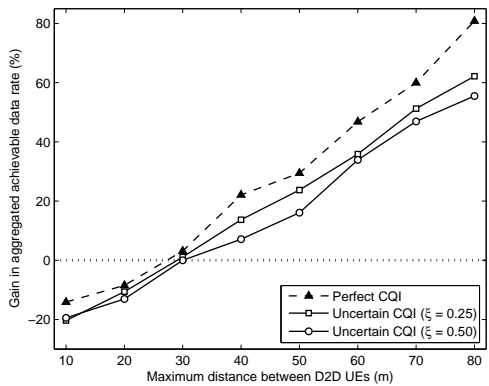}
\caption{Gain in aggregate achievable data rate for both perfect and uncertain CQI parameters. For uncertain CQI, uncertainty bound $\xi = 0.25$ and $\xi = 0.50$ mean that uncertainty in CQI parameters is not more than $25\%$ and $50\%$, respectively. For both the perfect and uncertain cases, there is a critical distance beyond which relaying of D2D traffic provides significant performance gain.}
\label{fig:gain_vs_d2d_dist}
\end{figure}
\begin{figure}[h t b]
\centering
\includegraphics[scale=0.50]{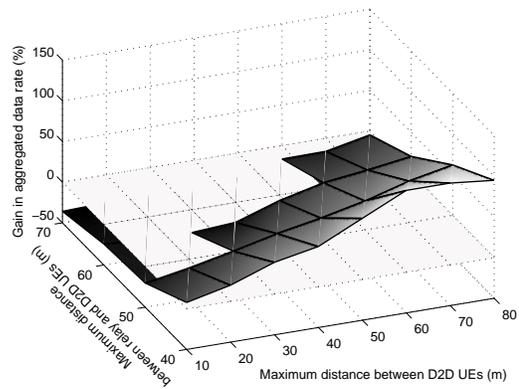}
\caption{Effect of relay distance on rate gain: $|\mathcal{C}| = 15$,  $|\mathcal{D}| = 9$. Uncertainty in CQI parameters is assumed to be not more than $25\%$. For every $D_{r,d}$, there is a distance threshold (i.e., upper position of the light-shaded surface) beyond which relaying provides significant gain in terms of aggregate achievable rate.} 
\label{fig:rel_var_sm}
\end{figure}

\begin{enumerate}[(i)]
\item \textit{Average achievable data rate vs. distance between D2D UEs:} The average achievable data rates of D2D UEs for both the proposed and reference schemes are illustrated in Fig. \ref{fig:rate_vs_d2d_dist}. Although the reference scheme outperforms when the distance between the D2D UEs is small (i.e., $d < 40~ \text{m}$), our proposed approach, which uses relays for D2D traffic, can greatly improve the data rate especially when the distance increases. This is due to the fact that when the distance increases, the performance of direct communication deteriorates due to increased signal attenuation. Besides, when the D2D UEs share resources with only one CUE, the spectrum may not be utilized efficiently, and therefore, the achievable rate decreases. As a result, the gap between the achievable rate with our proposed algorithm and that with the reference scheme widens when the distance increases.

\item \textit{Gain in aggregate achievable data rate vs. varying distance between D2D UEs:} The gain in terms of aggregate achievable data rate under both uncertain and perfect CQI is shown in Fig. \ref{fig:gain_vs_d2d_dist}. We calculate the rate gain as follows: 
$R_{\rm gain} = \frac{R_{\rm prop} - R_{\rm ref}}{R_{\rm ref}} \times 100 \%$,
where $R_{\rm prop}$ and  $R_{\rm ref}$ denote the aggregate data rate for the D2D UEs in the proposed scheme and the reference scheme, respectively. The figure shows that, compared to direct communication, with the increasing distance between D2D UEs, relaying provides considerable gain in terms of achievable data rate and hence spectrum utilization. As expected, the gain reduces under channel uncertainty since the algorithm becomes cautious against channel fluctuations and allocates RBs and power accordingly to protect the receiving nodes in the network. Note that there is a trade-off between performance gain and robustness against channel uncertainty. For example, when the distance $D_{r,d} = 50 ~\text{m}$, the performance gain of relaying under perfect CQI is $30\%$. In the case of uncertain CQI, the gain reduces to $24\%$ and $16\%$ for the uncertainty bound parameter $\xi = 0.25$ and $\xi = 0.50$, respectively. As the uncertainty bounds increase, the system becomes more roust against uncertainty; however, the achievable data rate degrades.

\item \textit{Effect of relay-UE distance and distance between D2D UEs on rate gain:} The performance gain in terms of the achievable aggregate data rate under different relay-D2D UE distances is shown in Fig. \ref{fig:rel_var_sm}. It is clear from the figure that, even for relatively large relay-D2D UE distances, e.g., $D_{r,d} > 60 ~\text{m}$, relaying D2D traffic provides considerable rate gain for distant D2D UEs.

\end{enumerate}

\section{Conclusion} \label{sec:conclusion_sm}

We have investigated the radio resource allocation problem in a relay-aided D2D network considering uncertainties in wireless channels and provided an iterative distributed solution to this problem using stable matching. We have analyzed the stability, uniqueness, and optimality of the proposed solution. We have also analyzed the complexity of the proposed approach. Numerical results have shown that the distributed solution is  close to the centralized optimal solution  with significantly lower computational complexity. We have compared the proposed relay-aided D2D communication scheme with an underlay D2D communication scheme. As an extension of this work, design and analysis of a unified mdium access control (MAC) protocol incorporating mode selection, device discovery, and such relay-aided D2D communication in the context of LTE-A network will be worth investigating.

\appendix
\numberwithin{equation}{section} 
\setcounter{equation}{0}  

\subsection{Proof of Proposition \ref{prop:stability}}
\label{appsec:stability}

Note that any arbitrary matching is not necessarily stable. 
In the following, we show that for any given preference profiles, each iteration of \textbf{Algorithm \ref{alg:joint_alloc}} ends up with a stable matching (i.e., there is no blocking pair). We prove the proposition by contradiction. Let $\mu_l$ be a matching obtained by \textbf{Algorithm \ref{alg:rb_alloc}} at any step $t$ of \textbf{Algorithm \ref{alg:joint_alloc}}. Let us assume that RB $n$ is not allocated to UE $u_l$, but it has a higher order in the preference list. According to this assumption, the $(u_l, n)$ pair will block $\mu_l$. 

Since the position of $u_l$ in the preference profile of $n$ is higher compared to the user $\hat{u}_l$ that is matched by $\mu_l$, i.e., $u_l \succeq \mu_l(n)$, RB $n$ must select $u_l$ before the algorithm terminates. However,  the pair $(u_l, n)$ does not match each other in the matching outcome $\mu_l$. This implies that $u_l$ rejects $n$ (e.g., line $7$ in \textbf{Algorithm \ref{alg:rb_alloc}}) and $(\hat{u}_l, n)$ is a better assignment. As a result, the pair $(u_l, n)$ will not block $\mu_l$, which contradicts our assumption. Consequently, the matching outcome $\mu_l$ leads to a stable matching since no blocking pair exists and the proof concludes.

\subsection{Proof of Proposition \ref{prop:uniq}} \label{appsec:uniq}

The proof is followed by the induction of number of users $U_l$, that are supported by relay $l$. For instance, let $\kappa_{u_l} = 1, ~\forall u_l \in \mathcal{U}_l$ (the proof for $\kappa_{u_l} > 1$ can be done analogously introducing dummy rows [i.e., UEs] in the utility matrix). The basis (i.e., $U_l = 1$) is trivial, since the only user definitely gets the best RBs according to her preference. When $U_l \geq 2$, let us consider $R_{i}^{(j)}$ to be the maximal entity of the utility matrix  $\mathfrak{U}_l$. For instance, let the matrix $\widehat{\mathfrak{U}}_l$ be obtained by removing the $i$-th row and $j$-th column from the utility matrix $\mathfrak{U}_l$. If $\mu_l$ is a stable matching for $\mathfrak{U}_l$, then by definition $\mu_l(i) = j$ and hence $\mu_l \setminus \lbrace (i,j) \rbrace$ must be a stable matching for $\widehat{\mathfrak{U}}_l$. By induction, there exists a unique stable matching $\widehat{\mu}_l$ for the smaller matrix $\widehat{\mathfrak{U}}_l$. Therefore, the proof is concluded due to the fact that $\mu_l = \widehat{\mu}_l \cup \lbrace (i,j) \rbrace$ is the unique stable matching for the utility matrix $\mathfrak{U}_l$.

\subsection{Proof of Proposition \ref{prop:optimality}}
\label{appsec:optimality}

Without loss of generality, let $\mathfrak{R}_{u_l, l}(\mu_l)$ denote the data rate achieved by UE $u_l$ for any matching $\mu_l$  for given uncertainty bounds and $\mathfrak{R}_l(\mu_l) = \displaystyle \sum_{u_l \in \mathcal{U}_l} \mathfrak{R}_{u_l, l}(\mu_l)$ is the sum-rate of all UEs. On the contrary,  let $\widehat{\mu}_l$ denote an arbitrary unstable outcome better than $\mu_l$, i.e., $\widehat{\mu}_l$ can achieve a better sum-rate. There are two cases that make $\widehat{\mu}_l$ unstable: \begin{inparaenum}[\itshape 1\upshape)]
\item lack of individual rationality, and/or
\item blocked by a UE-RB pair \cite{sm_def}.
\end{inparaenum} We analyze both the cases below.

\vspace{0.2cm}
\textit{Case 1 (lack of individual rationality):} If RB $n$ is not individually rational, then the utility of $n$ can be improved by removing user $\widehat{\mu}_l(n)$ with any arbitrarily user $u_l = \mu_l(n)$. Hence, the utility of $u_l$ increases and $ \mathfrak{R}_{u_l, l}(\widehat{\mu}_l) <  \mathfrak{R}_{u_l, l}(\mu_l)$.

\textit{Case 2 ($\widehat{\mu}_l$ is blocked):} When $\widehat{\mu}_l$ is blocked by any UE-RB pair $(u_l, n)$, RB $n$ strictly prefers UE $u_l$  to $\widehat{\mu}_l(n)$ and one of the following conditions must be true:

\textit{(i)} $u_l$ strictly prefers $n$ to some $\hat{n} \in \widehat{\mu}_l(u_l)$, or

\vspace{0.2cm}
\textit{(ii)} $\overline{\overline{\widehat{\mu}_l(u_l)}} < \kappa_{u_l}$ and $n$ is acceptable to $u_l$.

If  condition \textit{(i)} is true, we can obtain a stable matching $\mu_l$ by interchanging $n$ and $\hat{n}$ for $u_l$ as follows:
\begin{equation}
\mu_l(u_l) = \left\lbrace \widehat{\mu}_l(u_l) \setminus \hat{n} \right\rbrace \cup n.
\end{equation}
Hence, the new data rate of UE $u_l$ is 
\begin{align}
\mathfrak{R}_{u_l, l}(\mu_l) &= \sum_{j \in \mu_l(u_l)} R_{u_l}^{(j)} = R_{u_l}^{(n)} + \sum_{  \substack{j \in \mu_l(u_l), \\ j \neq n }} R_{u_l}^{(j)} \nonumber \\
&> R_{u_l}^{(\hat{n})} + \sum_{  \substack{j \in \widehat{\mu}_l(u_l), \\ j \neq n }} R_{u_l}^{(j)} =  \sum_{  j \in \widehat{\mu}_l(u_l)} R_{u_l}^{(j)} =   \mathfrak{R}_{u_l, l}(\widehat{\mu}_l)
\end{align}
where  $R_{u_l}^{(n)}$ is given by (\ref{eq:rate_sm_uncrt}). Since $u_l$ strictly prefers RB $n$ to $\hat{n}$ and the data rates for other UEs remain unchanged, for condition \textit{(i)}, it can be shown that $\mathfrak{R}_{l}(\mu_l) \geq \mathfrak{R}_{l}(\widehat{\mu}_l)$. 

\vspace{0.2cm}
When condition \textit{(ii)} is true, 
\begin{align} \label{eq:sm_case21}
\mathfrak{R}_{u_l, l}(\mu_l) &= \sum_{j \in \widehat{\mu}_l(u_l)} R_{u_l}^{(j)} + R_{u_l}^{(n)} \nonumber \\
&> \sum_{j \in \widehat{\mu}_l(u_l)} R_{u_l}^{(j)} = \mathfrak{R}_{u_l, l}(\widehat{\mu}_l).
\end{align}
Let $\hat{u}_l = \widehat{\mu}_l(n)$ with data rate $R_{\hat{u}_l}^{(n)}$. Then
\begin{align} \label{eq:sm_case22}
\mathfrak{R}_{\hat{u}_l, l}(\mu_l) &= \sum_{j \in \widehat{\mu}_l(\hat{u}_l)} R_{\hat{u}_l}^{(j)} - R_{\hat{u}_l}^{(n)} \nonumber \\
&< \sum_{j \in \widehat{\mu}_l(\hat{u}_l)} R_{\hat{u}_l}^{(j)} = \mathfrak{R}_{\hat{u}_l, l}(\widehat{\mu}_l).
\end{align}
From (\ref{eq:sm_case21}) and (\ref{eq:sm_case22}), neither $\mathfrak{R}_{l}(\mu_l) > \mathfrak{R}_{l}(\widehat{\mu}_l)$ nor $\mathfrak{R}_{l}(\widehat{\mu}_l) > \mathfrak{R}_{l}(\mu_l)$. Since for both cases \textit{1\upshape)} and \textit{2\upshape)} there is no outcome $\widehat{\mu}_l$ better than $\mu_l$, by \textbf{Definition \ref{def:pareto}},  $\mu_l$ is an optimal allocation and the proof follows.

\bibliographystyle{IEEEtran}

\end{document}